 \documentclass[final,5p,times,twocolumn,authoryear]{elsarticle}
 
\usepackage{amssymb}

\usepackage{color}
\usepackage{multirow,array}
\usepackage{url}

 \journal{Elsevier journal}

\begin{document}

\begin{frontmatter}



\title{Inference of Monosynaptic Connections from Parallel Spike Trains: A Review}


\author[Tokyo1,Tokyo2]{Ryota Kobayashi}
\ead{r-koba@k.u-tokyo.ac.jp}
\affiliation[Tokyo1]{organization={Graduate School of Frontier Sciences, The University of Tokyo},   
city={Chiba}, postcode={277-8561}, 
country={Japan}}
\affiliation[Tokyo2]{organization={Mathematics and Informatics Center, The University of Tokyo},   
city={Tokyo}, postcode={113-8656}, 
country={Japan}}

\author[Kyoto,Rits]{Shigeru Shinomoto}
\affiliation[Kyoto]{organization={Graduate School of Informatics, Kyoto University},   
city={Kyoto}, postcode={606-8501}, 
country={Japan}}
\affiliation[Rits]{organization={Research Organization of Science and Technology, Ritsumeikan University},   
city={Shiga}, postcode={525-8577}, 
country={Japan}}

\begin{abstract}
This article presents a mini-review about the progress in inferring monosynaptic connections from spike trains of multiple neurons over the past twenty years.  
First, we explain a variety of meanings of ``neuronal connectivity'' in different research areas of neuroscience, such as structural connectivity, monosynaptic connectivity, and functional connectivity. Among these, we focus on the methods used to infer the  monosynaptic connectivity from spike data. 
We then summarize the inference methods based on two main approaches, i.e., correlation-based and model-based approaches. Finally, we describe available source codes for connectivity inference and future challenges.  
Although inference will never be perfect, the accuracy of identifying the monosynaptic connections has improved dramatically in recent years due to continuous efforts. 
\end{abstract}




\end{frontmatter}


%
%
\section{Introduction}

It has been about a hundred years since it was discovered that neurons generate electrical pulses and that these electrical pulses carry information between neurons. As it became possible to chronically measure the activity of a neuron during animal behavior, scientists began to investigate how individual neurons are involved in a neural response to stimuli. In recent years, it has become possible to record spike signals from multiple neurons in parallel, and the number of such neurons has increased over the years, now exceeding a thousand~\citep{Stevenson2011a, Steinmetz2021}. Although this number is still small compared to the total number of neurons in the brain, it is likely to increase in the future, and this information will eventually yield a great deal of knowledge about neural information processing. 

A straightforward way to study the function of the animal's nervous system is to examine how stimuli are represented in neural activity in the brain and how this neural activity regulates the animal's behavior. For example, neuronal representation in the visual system has been analyzed by measuring individual cells in the visual cortex and exploring the visual stimuli to which these cells respond~\citep{hubel1977ferrier, fujita1992columns}; pioneering studies of the motor control of animal behavior have revealed the role of individual neurons in operantly conditioned movements by recording neuronal activity in the sensorimotor cortex of animals during behavior~\citep{evarts1968relation}. 
These early studies considered a single neuron as representative of the myriad of cells in each brain region that work for a single function. 
Recent advances in measurement technology have made it possible to record the activity of many individual cells in parallel and to elucidate the role of multiple neurons in animal behavior. 
Assuming independence among neurons, information about the stimulus-response relationship of individual cells increases in proportion to the number of cells measured simultaneously. 

In addition, recent measurement technology not only allows us to obtain information about the correlation between stimulus and single cell activity, but also provides information about the correlation of activity between neurons, allowing us to capture information transmission at the level of individual neurons. 
The amount of information about correlations between neurons increases  in proportion to the square of the number of cells measured, and it increases faster than the amount of information about the stimulus-response relationship. 
In 1967, more than half a century ago, when it was barely possible to measure the activity of multiple neurons simultaneously, Perkel, Gerstein, and Moore proposed a method for inferring monosynaptic connections between neurons by evaluating the interdependence of multiple neuronal spike trains~\citep{Perkel1967}. 

However, neurons in the brain are densely interconnected and influence each other in complex ways. There is often interdependence in the firing of neurons even when they are not directly connected, and this classical method  has been shown  to make many false inferences. 
In this article, we review the advances in the methods for inferring monosynaptic connections from spike data over the last twenty years. Finally, we discuss the limitations and future challenges of current methods.

%
%
\begin{figure*}[h]
\centering
\includegraphics[width=14cm]{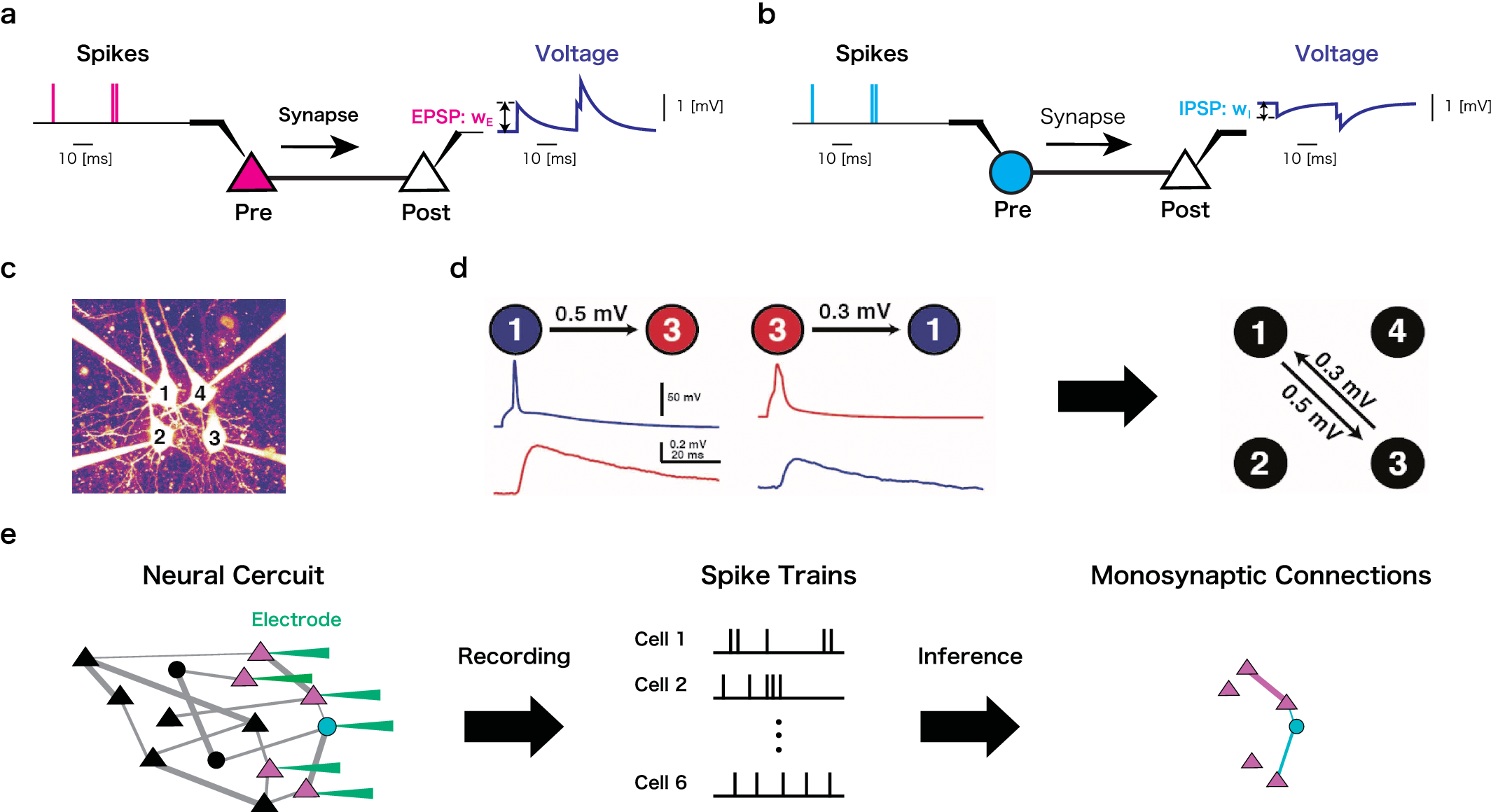} 
\caption{Monosynaptic connections between neurons. 
(a) and (b): Signal transmission between neurons. Due to the synaptic connection, the membrane potential of a postsynaptic neuron (Post) increases (a) or decreases (b) after an action potential of a presynaptic neuron (Pre).  Whether the membrane potential increases or decreases depends on the type of presynaptic neuron. 
While an excitatory neuron (triangle) increases the membrane potential of the postsynaptic neuron, an inhibitory neuron (circle) decreases it. The strength of the monosynaptic connection is defined as the change in potential (EPSP or IPSP). 
(c) and (d): Experimental measurement of the monosynaptic connections. Using quadruple whole-cell recordings (c), the EPSP or IPSP can be measured by evoking action potentials in each of the four neurons and measuring the change in the voltage in the other neurons (d). Panels (c) and (d) are adapted and modified with permission from \cite{Song2005}. 
(e): Inference of monosynaptic connections from spike trains of multiple neurons. The monosynaptic connections can be inferred from the correlation between spike trains. In this review, we focus on the advances in the method to infer/estimate the monosynaptic connections from spike data.  
 }
\label{fig:Est_Problem}
\end{figure*} 
\newpage
\section{Variety of meanings for ``neuronal connectivity''}

The nervous system in the brain consists of a myriad of neurons that transmit signals through synaptic connections to produce the neural response of an animal. 
Bottom-up research on brain function analyzes the circuit structure of how neurons are connected via synapses and elucidates the mechanisms by which these connections are plastically modified. Before getting to the main topic, however, we must first mention that the term ``connectivity'' or ``connection'' is used in three ways depending on the field of neuroscience, i.e., structural connectivity, monosynaptic connectivity, and functional connectivity.

\subsection{Structural connectivity}
Structural connectivity means how neurons are anatomically connected.  
Studies that attempt to analyze the structure of neural circuits include efforts to reconstruct neurons in three-dimensional space using electron microscopy~\citep{Helmstaedter2013, Scheffer2020}. 
To study the structural connections between distant neurons or between brain regions, connections are first visualized at the neuronal level using horseradish peroxidase and other techniques~\citep{Binzegger2004}. Structural connectivity at the macroscopic level between brain regions is analyzed using diffusion tensor imaging (DTI) that visualizes the bundles of nerve fibers. The structural connections obtained in this way have been used to simulate neural circuits to discuss the function of columns and the impact of a single neuronal spike on the dynamics at the level of the whole brain~\citep{Izhikevich2008, schmidt2018multi}.

\subsection{Monosynaptic connectivity} 

The mechanism of signal propagation between neurons is that action potentials (spikes) generated in one neuronal cell (presynaptic neuron) propagate through axons and, upon arrival at synapses at the axon terminals, chemical transmitters are released, causing changes in the membrane potential of a postsynaptic neuron (Fig.\ref{fig:Est_Problem}a, b). When the membrane potential of a neuron reaches the threshold, the neuron generates a spike. 
Monosynaptic connectivity is defined as a direct influence between neurons in their membrane potential (i.e., EPSP or IPSP).    
To confirm the monosynaptic connections, we can stimulate one neuron to generate an action potential and observe whether the activation of the neuron results in a change in the membrane potential of another neuron (Fig.\ref{fig:Est_Problem}c, d). 
If this measurement can be done, not only the presence or absence of synaptic connections, but also the strength of the connections can be measured in units of excitatory/inhibitory postsynaptic potentials (EPSP/IPSP)~\citep{Thomson2002, Song2005}. 

Although synaptic connections can be rigorously determined by intracellular recording, intracellular recording in behaving animals is challenging, and the number of cells that can be measured simultaneously is limited to a few, even for brain slices ({\it in vitro}). In contrast, technological innovations are continuously increasing the number of cells that can be measured simultaneously using extracellular recording. 
The presence of a monosynaptic connection can be inferred from the correlation between spike trains with a time difference of a few milliseconds (Fig.\ref{fig:Est_Problem}e). Although the inference is subject to some error, it may be possible to improve the accuracy of the inference by improving the data analysis method. 
In this paper, we review the methods that have been developed for this purpose.

\subsection{Functional connectivity} 

Neurons in the brain are connected at multiple levels, even if they are not connected monosynaptically, and their neural activity may not be independent of each other. 
Many studies have attempted to detect correlations of activity between different brain regions, particularly in humans, using non-invasive measurements such as fMRI,  in search of potential features associated with neurological and psychiatric disorders~\citep{Lynall2010, Van2019}. 
They refer to the correlation as ``functional connectivity'', but this simply means an apparent correlation between brain regions or between neurons, and it ``does not provide direct insight into how these correlations are mediated''~\citep{Friston1994}. It should be noted that the existence of a strong functional connection (direct or indirect) between two given elements cannot be attributed to the physical connection between the two elements. 

%
%
\section{Methods for inferring monosynaptic connections} 

This review focuses on studies that aim to infer or estimate direct monosynaptic connections between neurons. We will refer to the studies we focus on as ``inferring monosynaptic connections'' to distinguish them from studies that simply detect activity correlations as ``functional connections''. 

\begin{figure}[h]
\centering
\includegraphics[width=7cm]{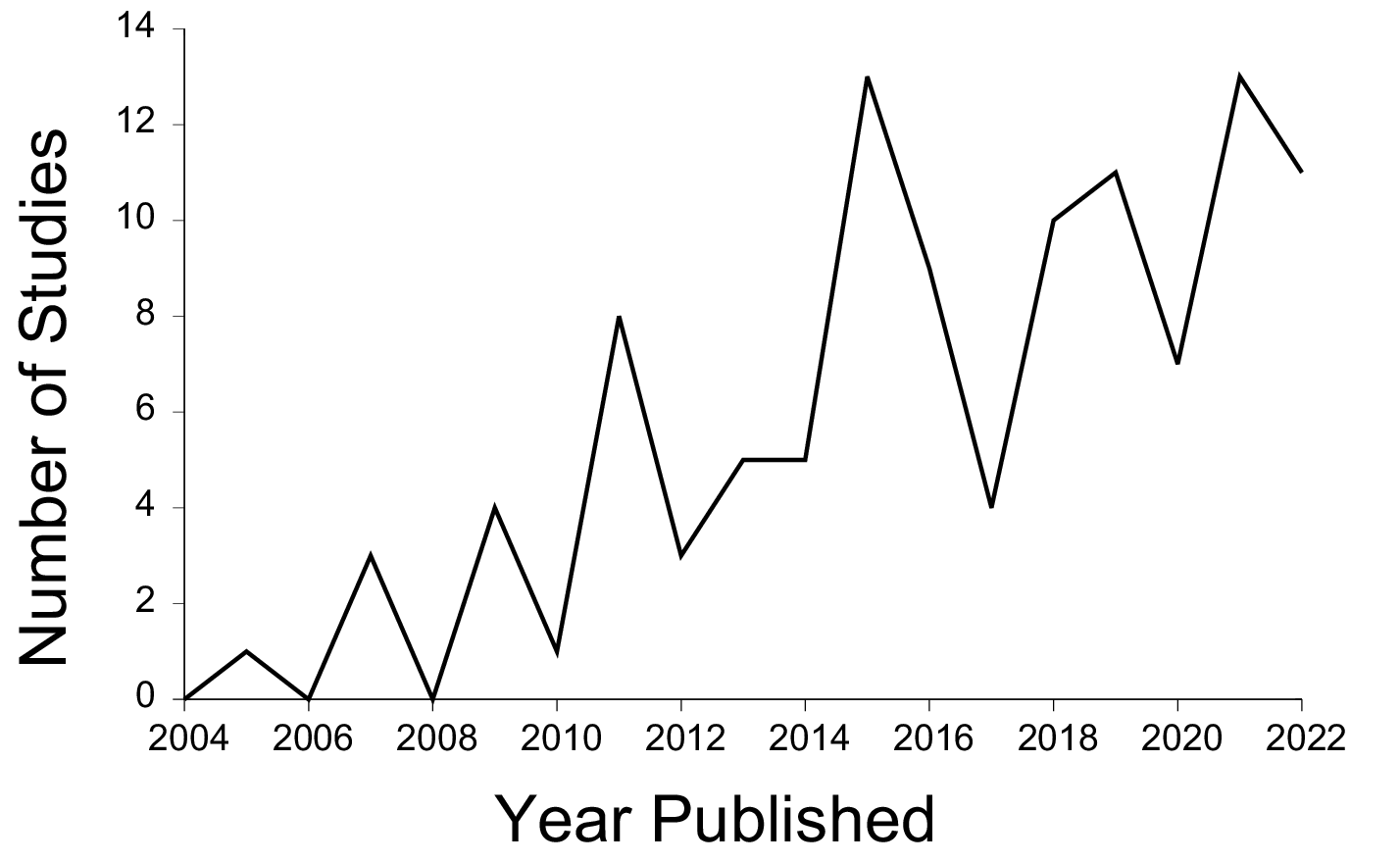} 
\caption{Number of papers that contain keywords of ``connectivity, estimation, spike," ``connectivity, inference, spike," or ``connectivity, reconstruction, spike."}
\label{fig:Num_Paper}
\end{figure}

The number of papers aiming at the inference of monosynaptic connectivity has increased in recent years (Fig.~\ref{fig:Num_Paper}). We searched Web of Science for papers containing the keywords ``connectivity, inference, spike,'' ``connectivity, estimation, spike,''  or ``connectivity, reconstruction, spike.'' in the title or abstract. The search returned 308 articles. We found 124 articles that addressed the problem of inferring monosynaptic connections between neurons. Even excluding the 8 articles that used intracellular recording to measure synaptic connectivity and one article that focused on a graphical user interface (GUI) to analyze recorded data, we still have 115 articles.

Methods for inferring the monosynaptic connections between neurons can be divided into two categories: 
1) methods based on the statistical characteristics of the spike correlation between neurons and 
2) methods based on the model of the dynamic spike generation process of the recorded cells. 
In this paper, we call them ``correlation-based'' and ``model-based'', respectively  (Table~\ref{tab:method}). The former has sometimes been called ``model-free''~\citep{Stetter2012,Vicente2011, de2018connectivity}. Recently, however, it has been realized that applying a statistical model such as the generalized linear model to the spike correlation is an efficient way to analyze the data (e.g.~\cite{Kobayashi2019}), and we have changed the category name accordingly. 
In addition, information-theoretic quantities, such as mutual information, are measures of correlation, because they evaluate the dependence between two random variables~\citep{Cover1991}.  
In the following sections, we will further subdivide them with brief methodological explanations, and finally introduce open source codes that are currently available.

\begin{table}[t!]
\centering
\caption{Approaches to infer monosynaptic connections.} \label{tab:method} 
\vspace{0.5cm}
\begin{tabular}{|c|c|c|}
\hline 
Category & Method & Papers ( \%) \\ \hline 
\multirow{3}*{Correlation-based} & Cross-correlation & 26 (23 \%) \\
& Information theory & 11 (10 \%) \\ 
& Others & 24 (21 \%) \\ \hline 
\multirow{2}*{Model-based} & GLM & 36 (31 \%) \\	  
& Others & 18 (16 \%) \\ \hline
\end{tabular}
\end{table} 

%
%
\subsection{Correlation-based approach}

Methods based on the spike-occurrence correlation of two cells occupy 53\% of all studies of monosynaptic connection estimation. Here we further subdivide them by the terms ``cross-correlation,'' ``information theory,'' and ``others'' (Table~\ref{tab:method}).

\subsubsection{Cross-correlation method}

The membrane potential of a neuron fluctuates due to the input from other neurons transmitted via synaptic connections.  A spike arriving at an excitatory/inhibitory synapse increases/decreases the membrane potential of a target neuron, i.e., the postsynaptic neuron (Fig.1a, b). Accordingly, the probability of a spike occurring at the postsynaptic neuron (Post in Fig.1a, b) increases/decreases for a few milliseconds after the occurrence of a spike at another neuron (Pre in Fig.1a, b).

\begin{figure*}[t]
\centering
\includegraphics[width=15cm]{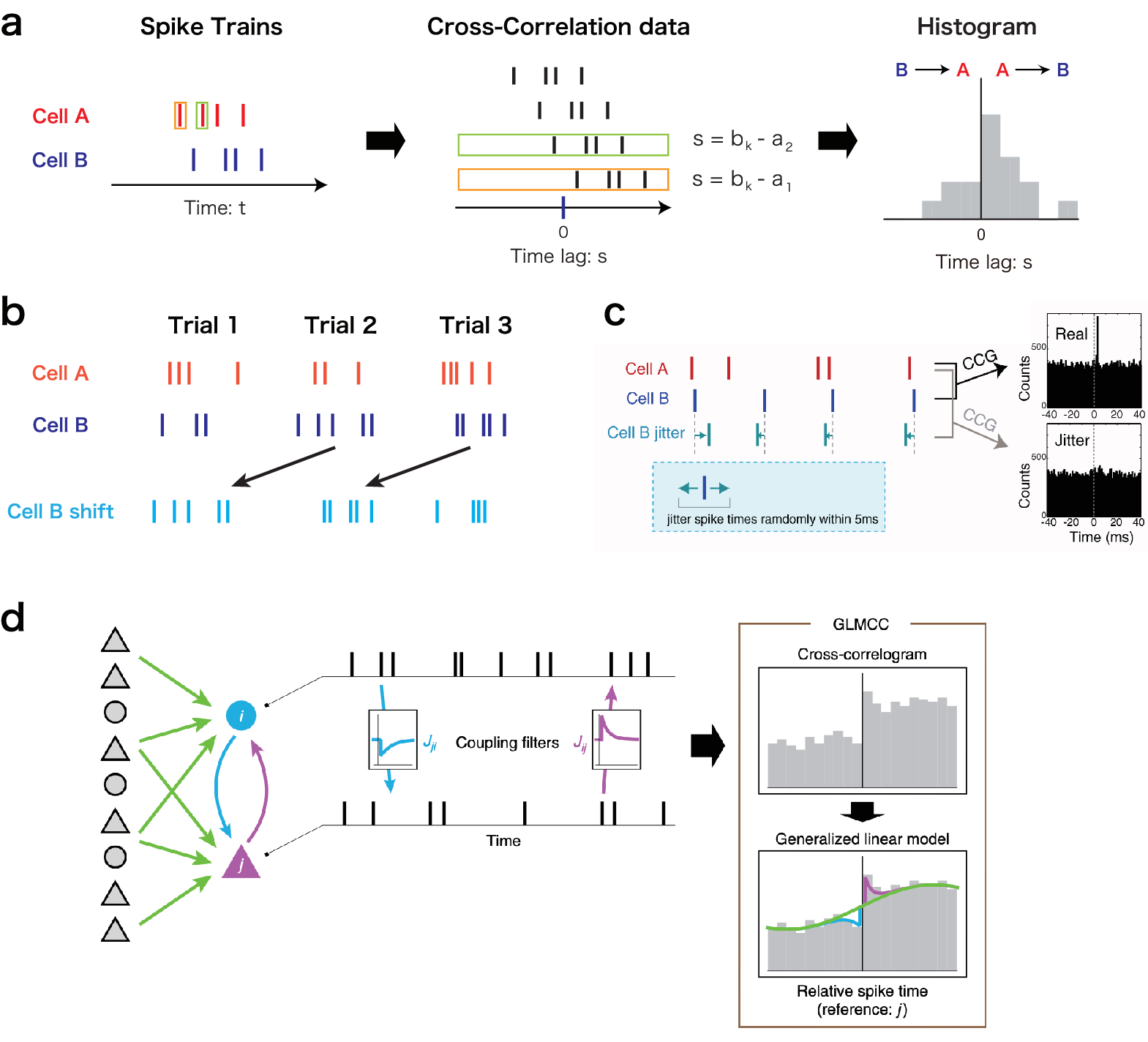} \newline
\caption{
Classical cross-correlation method and its extensions. 
(a): Classical cross-correlation method. First, the cross-correlation data is obtained by computing the time lag $s= b_k- a_j$ relative to the reference cell (cell A). Note that $a_k$ ($b_k$) is the $k$ th spike time of cell A (B).
A histogram of the CC data (CCG) is then constructed. If the CCG has a peak or dip in the range of time difference $s> 0$ ($s<0$), it suggests a synaptic connection from cell A to B (cell B to A). 
(b): Shift predictor.  First, an artificial spike train (Cell B shift: cyan) is generated by shifting the trial of the target cell (Cell B: blue). The shift predictor is then obtained by calculating the CCG between Cell A (red) and Cell B shift (cyan). Finally, the connection is inferred from the histogram obtained by subtracting the shift predictor from the original CCG between cell A (red) and cell B (blue). 
(c): Jittering. A jittered spike train (Jitter Cell B: cyan) is generated by randomly ($\pm$ 5 [ms]) shifting the spike of the target cell (Cell B: blue). Monosynaptic connectivity is inferred by comparing the CCG  between Cell A (red) and Cell B (blue) with 
the CCG  between Cell A (red) and Jitter Cell B (cyan). Panel (c) is adapted with permission from \cite{Fujisawa2008}. 
(d): Decomposition. This approach infers the synaptic connectivity by decomposing the CCG into multiple components. 
For example, GLMCC estimates synaptic connections by decomposing the CCG into the effects from cells other than the two cells (green) and the effects due to synaptic connections between the two cells (cyan, magenta). Panel (d) is adapted with permission from \cite{Kobayashi2019}. 
}
\label{fig:CC}
\end{figure*}

The classical cross-correlation (classical CC) method attempts to find the existence of synaptic connections between neurons by analyzing the cross-correlation of spike occurrences of two neurons to detect the causal influence due to the synaptic connections~\citep{Perkel1967,Moore1970}. The method consists of the following three steps (Fig.\ref{fig:CC}a). 

\begin{enumerate}

\item Obtain a cross-correlation by computing the time of spikes of a target neuron (cell B) measured relative to each spike of the reference neuron (cell A) (Fig.\ref{fig:CC}a: center).

\item Construct the cross-correlation histogram or cross-correlogram (CCG) (Fig.\ref{fig:CC}a: right).

\item Determine that there may be a synaptic connection from A to B (or B to A) if the CCG has a statistically significant peak or dip  in the region $s>0$ (or $s<0$).

\end{enumerate}

\clearpage

There are several variants of the classical method: the spike coherence~\citep{Brillinger1976}, the joint peri stimulus time histogram (JPSTH)~\citep{Gerstein1969, Aertsen1989}, and the covariance~\citep{Pernice2013}.

The classical CC method has the advantage of high interpretability because the influence of the synaptic connection is visible from the CCG. However, the CCGs often exhibit large fluctuations, which may be caused by a large number of other neurons that are not recorded, and such large fluctuations may induce many false detections. To overcome this difficulty, there have been several approaches to improve the estimation. In the following, we introduce four main approaches to extend the classical CC method.

\begin{itemize}

\item {\bf Shift predictor (Fig.~\ref{fig:CC}b):} 
A CCG constructed from two neurons often exhibits a broad peak near the zero time lag, even when the neurons are not monosynaptically connected. A shift predictor has been proposed to avoid the false inference caused by this peak~\citep{Perkel1967, Toyama1981, Palm1988}. 
When an animal is repeatedly exposed to the same stimulus, this activation causes a broad peak in the CCG. 
The correlation due to the stimulus may be extracted by constructing a CCG between a reference cell (Cell A) and an artificial spike train (Cell B shift) obtained by shifting the response of a trial, which is called the shift predictor. 
The influence of a monosynaptic connection is expected to appear in the difference between the original CCG and the shift predictor. 

\item {\bf Jittering (Fig.~\ref{fig:CC}c):}
While a CCG may also exhibit slow fluctuations caused by background activity, the monosynaptic effect is expected to show up as a sharp peak or dip typically in a few milliseconds near the origin ($s= 0$). The sharp peak or dip near the origin may be destroyed if the original spike times are jittered for several milliseconds, while the slow fluctuations remain. 
The jittering method estimates the statistical significance of a peak or a dip in the original CCG by constructing many CCGs from jittered spike trains~\citep{Fujisawa2008, Amarasingham2012}.

\item {\bf Decomposition (Fig.~\ref{fig:CC}d):} 
This approach extracts the effect of the monosynaptic connections by decomposing the CCG into multiple components. 
\cite{Kobayashi2019} applied the generalized linear model to decompose the CCG into monosynaptic influences and background fluctuations caused by other neurons. The method called GLMCC was able to estimate synaptic connections more accurately than the classical CC and Jittering when applied to synthetic spike trains generated by simulating a network of 1,000 Hodgkin-Huxley model neurons and a network of 10,000 leaky integrate-and-fire neurons. \cite{Ren2020} employed a similar model and showed that the estimation accuracy can be improved by taking the neuron type into account. \cite{Spivak2022} proposed to deconvolute the CCG into the effect of spiking of a reference neuron, a target neuron, and other neurons. More recently, \cite{tsubo2023non} pointed out that a cusp often appears at the origin of many biological CCGs, and proposed a new method, ShinGLMCC, to avoid making false inferences caused by the cusp.

\item {\bf Post-processing:}
CCGs of {\it in vivo} data are often jagged, burying a clear peak or dip. There are several suggestions for post-processing to find a peak/dip in a CCG. 
\cite{Pastore2018} suggested subtracting the filtered mean from the original CCG. 
\cite{De2019} used an edge filter used in image processing and processed the CCG. They generated synthetic data by simulating a network of 1,000 Izhikevich model neurons and showed that their method provided better estimation.
\cite{Endo2021} trained the convolutional neural network with synthetic data from 1,000 MAT model neurons~\citep{Kobayashi2009b} to learn the synaptic connectivity. 

\end{itemize} 

\subsubsection*{Application to experimental data}
The classical CC method and its extensions are not only validated with synthetic data, but also applied to many experimental data.

\begin{itemize}
\item {\bf Classical Cross-Correlation:} 
\cite{Bartho2004} used the classical CC method to estimate synaptic connections between neurons in layer 5 of the somatosensory cortex and the prefrontal cortex of a rat. The number of connections detected was 0.2 percent of the 60,000 pairs analyzed. The estimation accuracy of the classical CC method was evaluated using synthetic data obtained from a network of 60 Izhikevich neurons~\citep{Garofalo2009} or a network of 5,000 Hodgkin-Huxley neurons~\citep{Kobayashi2013}. 
\item {\bf Shift predictor:} 
Shift predictor has been used to analyze many experimental studies: Cat visual cortex~\citep{Toyama1981}, Cat thalamus to visual cortex~\citep{Reid1995}, Rat visual cortex~\citep{Yoshimura2005}, Rat thalamus to somatosensory cortex~\citep{Liew2021}, Monkey V1 and V2 area~\citep{Nowak1999}, and Monkey prefrontal cortex~\citep{Funahashi2000}.

\item {\bf Jittering:}
Jittering has been used to analyze rat medial prefrontal cortex~\citep{Fujisawa2008}, rat hippocampus~\citep{Mizuseki2009}, and rat prefrontal cortex~\citep{Schwindel2014}. 

\item {\bf Decomposition:} 
\cite{Kobayashi2019} proposed the GLMCC and applied it to hippocampal CA1 and EC~\citep{Mizuseki2013} measured from a freely behaving rat. The excitatory and inhibitory characteristics of individual neurons determined by GLMCC showed good agreement with those determined by experts based on spike waveforms and CCGs. Furthermore, the cell-to-cell binding probabilities were consistent with those previously identified in slice experiments.
\cite{Ren2020} proposed ExGLM and applied it to data measured from cultured cells {\it in vitro} using a large multielectrode array (512 electrodes). The presumed excitable characteristics were consistent with previous experimental results.
\cite{Spivak2022} applied their proposed method to mouse hippocampal CA1 data from three freely-moving mice recorded with high-density silicon probes. The results show that the proposed method can remove the effect of burst firing of presynaptic neurons.

\item {\bf Post-processing:}
\cite{Pastore2018} used their method, FNCCH, to estimate synaptic connections {\it in vitro} cell culture (4,096 units). The results suggest that the {\it in vitro} network has small-world, scale-free, and rich-club (strongly connected) properties. \cite{Endo2021} used their method, CoNNECT, to analyze data recorded from the prefrontal, IT, and V1 cortices of monkeys.

\end{itemize}
\subsubsection{Information-theoretic measure}

Information-theoretic measures such as joint entropy~\citep{Garofalo2009} or directed information~\citep{Quinn2011, So2012, Cai2017} have been used to estimate synaptic connections. 
Here we introduce the transfer entropy (TE)~\citep{Schreiber2000}, which is the most common among the 11 articles in Table~\ref{tab:method}, see \cite{Vicente2011} for a review. 

TE from one neuron $j$ to another neuron $i$ quantifies the degree to which the spike data of neuron $j$ can improve the prediction of the spike data of neuron $i$. If there is a strong excitatory synapse from neuron $j$ to neuron $i$, spike occurrence in neuron $j$ should have a strong impact on spike generation in neuron $i$, and accordingly the TE should take a large value. In contrast, if there is no synapse from neuron $j$ to neuron $i$, the activity of neuron $j$ should not affect the prediction of spike data of neuron $i$, and the TE is expected to vanish (${\rm TE}= 0$). 

When computing the TE, spike trains are discretized with a time bin $\Delta$, i.e. it is represented as a discrete time binary series, where $I_k=1$ ($J_k=1$) if the $i$ th ($j$ th) neuron generates spikes in the $k$ th time bin, otherwise $I_k=0$ ($J_k=0$).
The TE from one neuron $j$ to another neuron $i$ is calculated as
\begin{eqnarray}
	{\rm TE}_{ji} = \sum_{ I_k, I_{k-1}, J_{k-1} } p( I_k, I_{k-1}, J_{k-1} ) \log \frac{ p\left( I_k | I_{k-1}, J_{k-1} \right) }{p\left( I_k | I_{k-1} \right) },
\label{eq:def_TE}
\end{eqnarray} 
where $p\left( I_k | I_{k-1}, J_{k-1} \right)$ and $p\left( I_k | I_{k-1} \right)$ represent the conditional probabilities.
The advantage of the TE over the classical CC method is that it estimates synaptic connections taking into account the past activity of neuron $i$ itself, as can be seen from the equation (\ref{eq:def_TE}). This is expected to allow TE to estimate synaptic connections by incorporating neuronal properties such as refractory periods and bursts.

Improvements to the TE have also been proposed. For instance, \cite{Ito2011} proposed Higher Order TE (HOTE), which considers spiking data before $d$ steps $(I_{k-1}, I_{k-2}, \cdots I_{k-l}$; $J_{k-d}, J_{k-d-1}, \cdots J_{k-d-m})$ and showed it can improve the accuracy of estimation. 
Furthermore, the estimated TE value depends critically on the time bin $\Delta$, which potentially affects the inferred connectivity. 
For this reason, it has been proposed to fit GLM~\citep{Quinn2011, So2012} or to estimate the TE directly from interspike interval data~\citep{Shorten2021}. 

Information-theoretic measures have also been applied to experimental data. For example, the directed information has been used to estimate synaptic connectivity in monkey primary motor cortex~\citep{Quinn2011, So2012}. The extended TE has been used to estimate synaptic connectivity in slice cultures of rodent somatosensory cortex~\citep{Ito2011}. More recently, the TE has been used to examine changes in connections between neurons in rat embryo dissociated hippocampal neuron cultures {\it in vitro} at 6-35 days post-culture to determine how neural networks evolve as a result of maturation~\citep{Antonello2022}.

\subsubsection{Other methods} 

This section introduces correlation-based methods that are not based on the cross-correlation method or the information-theoretic measure. 

\begin{itemize}
\item {\bf Granger causality:}
Granger causality~\citep{Granger1969, Seth2007} has also been used to estimate monosynaptic connectivity. Since the original Granger causality analysis is proposed for time series, it needs to be modified for the analysis of point event data (point process) when considering neuronal spike data, e.g., by using the Fourier transform~\citep{Nedungadi2009}, introducing a likelihood ratio test for the point process model~\citep{Kim2011}, or simply smoothing the spike data~\citep{Shao2015}. 

\item {\bf Spike triggered non-negative matrix factorization (STNMF):}
A method combining spike-triggered average and non-negative matrix factorization (STNMF) has been proposed to estimate the synaptic connectivity in the retina ~\citep{Liu2017}.

\item {\bf Others:}
Other methods have also been proposed using the spike train metric~\citep{Kuroda2011} and the copulas~\citep{Sacerdote2012}.

\end{itemize}
\subsection{Model-based approach}
While the correlation-based approach directly analyzes the interdependence of two given spike trains, the model-based approach constructs a mathematical model in which spiking neurons interact. The model-based approach can take account of all recorded neurons simultaneously. The synaptic connections are determined by fitting a large number of model parameters to given spike trains. 
\subsubsection{Generalized linear model}
Many studies have used the generalized linear model (GLM)~\citep{Paninski2004, Okatan2005, Truccolo2005, Chen2022}, see~\cite{Stevenson2008a} for a review. The firing rate $\lambda_i(t)$ of the $i$ th neuron is given as a function of an input 
\begin{eqnarray}
	\lambda_i(t) = F( {\rm Input} ),
\end{eqnarray}
where the input is decomposed as
\begin{eqnarray}
		{\rm Input} &=& [{\rm Base}] + [{\rm Stimulus}] \nonumber \\
			&& \quad \quad + [{\rm Refractory}] + [{\rm Coupling}]. 
	\label{eq:GLM}
\end{eqnarray} 
The GLM (Eq.~\ref{eq:GLM}) assumes that neuronal firing is influenced by the following four factors. 
The first term [Base] represents the spontaneous firing rate. The second term [Stimulus] represents the effect of the stimulus and is often modeled by the convolution of the external stimulus and the temporal/spatial filter. 
The third term [Refractory] represents the property of neurons that are less likely to fire immediately after a spike, which is related to the refractory period or spike frequency adaptation~\citep{Koch2004, Kobayashi2016}. The fourth term [Coupling] represents the monosynaptic effect from other neurons recorded in parallel, and the synaptic connections are inferred from the parameters in this term. 

One of the reasons that GLM is often used is the applicability of the maximum likelihood method. The maximum likelihood method has been shown to provide more accurate parameter estimation than the method of moments~\citep{Brillinger1975} or other methods~\citep{Berry1998, Chornoboy1988, Paninski2004}.
For GLM, it has also been shown that the maximum likelihood method does not have a local maximum~\citep{Paninski2004}. The input-output functions $F$ can be the exponential function~\citep{Jolivet2006,Kobayashi2007} or the Relu function~\citep{Ermentrout1998,Kobayashi2009a}. GLMs whose input-output function is a Relu function are also called Hawkes process~\citep{Hawkes1971}, for which fast estimation algorithms have been developed~\citep{Lambert2018}. 

\begin{table*}[t!]
\centering
\caption{Available codes for connectivity inference from spike trains.} \label{tab:codes} 
\vspace{0.5cm} 
\small 
\begin{tabular}{|c|c|c|}
\hline 
Method & Language & URL \\ \hline 
TSPE~\citep{De2019} & MATLAB & \url{https://github.com/biomemsLAB/TSPE} \\ \hline 
\multirow{2}*{GLMCC~\citep{Kobayashi2019}} 
& Python & \url{https://github.com/NII-Kobayashi/GLMCC} \\
& Web application & \url{https://s-shinomoto.com/CONNECT} \\ \hline 
ExGLM~\citep{Ren2020} & MATLAB & \url{https://github.com/NaixinRen/extended-GLM-for-synapse-detection} \\ \hline 
\multirow{2}*{CoNNECT~\citep{Endo2021}} 
& Python & \url{https://github.com/shigerushinomoto/CoNNECT} \\
& Web application & \url{https://s-shinomoto.com/CONNECT} \\ \hline 
Transfer entropy~\citep{Ito2011} & MATLAB & \url{https://code.google.com/archive/p/transfer-entropy-toolbox} \\ \hline 
CoTETE~\citep{Shorten2021} & Julia / Python & \url{https://github.com/dpshorten/CoTETE.jl} \\ \hline 
GLM~\citep{Pillow2008} & MATLAB & \url{https://github.com/pillowlab/GLMspiketools} \\ \hline 
\end{tabular}
\end{table*} 

It should be noted that the number of parameters in the GLM tended to be too large and overfitting occurred when applied to experimental data~\citep{Stevenson2008a}. Research has been conducted to overcome this problem. One approach is to use L1 (Lasso) regularization to reduce the effective number of parameters~\citep{Paninski2004, Zaytsev2015, Lambert2018}. For example, ~\cite{Pillow2008} estimated significant functional connections between retinal ganglion cells by incorporating the physiological knowledge that synaptic connections are sparse using the Group Lasso method~\citep{Yuan2006}. Another approach is to reduce the effective number of parameters by Bayesian estimation~\citep{Gerwinn2010}. For example, ~\cite{Stevenson2008b} developed an estimation method that incorporates the smoothness of the postsynaptic potential and the sparsity of the connections into the prior distribution. \cite{Chen2011} reported that Hierarchical Bayesian modeling is more accurate than L1 regularization in estimating connections.

\subsubsection*{Applications to synthetic and experimental data}
Below we summarize the applications of GLMs to synthetic and experimental data.

\begin{itemize} 
\item {\bf Applications to synthetic data:} 
The validity of GLMs can be tested by estimating synaptic connectivity from spike data obtained by numerical simulation of an interconnected network of neurons. However, most GLM studies have estimated functional connectivity, and few studies have examined the correspondence between functional connectivity and actual synaptic connectivity (exceptions are \cite{Kobayashi2013} and \cite{Zaytsev2015}). 
\cite{Zaytsev2015} proposed a fast algorithm for estimating synaptic connections, and the proposed method was applied to simulated data based on a network of 1,000 spiking model neurons. It has been shown that the proposed method can accurately and quickly estimate synaptic connections even when 1,000 cells are measured simultaneously.

\item {\bf Applications to biological data:} 
An example of experimental validation of GLM is a study of the stomatogastric ganglion (STG) circuit in the crab ({\it Cancer borealis})~\citep{Gerhard2013}. In this study, the synaptic connections between neurons were estimated from spike data measured from the STG. The results showed that connections estimated using GLM were consistent with known physiological connections. In contrast, connections estimated using Granger causality and firing rate were not consistent with physiological connections.

GLM has also been applied to the retina~\citep{Pillow2008,Gerwinn2010}), hippocampus~\citep{Okatan2005,Xia2019}, and motor system~\citep{Stevenson2008b}. The neuroscientific significance of the inferred connections has also been investigated. For example, \cite{Pillow2008} showed that inferred connections between two types of cells (ON and OFF retinal ganglion cells) in the monkey retina improved the accuracy of predicting firing rates and decoding visual stimuli. This result suggests that the inferred connections may contribute to visual information processing in the retina.
In addition, some studies suggest a relationship between estimated connections and behavior. \cite{Xia2019} estimated the connection between the rat hippocampus (HPC) and prefrontal cortex (PFC). The results showed that the estimated connection corresponded to rat behavior, with strong connections between HPC and PFC observed only on trials where the rat responded correctly to the task. 

\end{itemize} 
\subsubsection{Other models}
In addition to the GLM or point-process models (Section 3.2.1),
other models have been used to infer monosynaptic connections from spike trains. We introduce two popular models for connectivity inference, namely the ising model and the spiking neuron model.

\begin{itemize} 
\item {\bf Ising model:}  
Ising model is a simple mathematical model of a magnetic material in which each spin can take either an up or down state. 
A memory retrieval process in a network of model neurons is analogous to an energy relaxation process in a network of Ising spins. The Ising model can be used to analyze a network of neurons interacting through synapses. This can be done by transforming the spike trains into a binary state, resting or firing, with the time bin $\Delta$. The connection between neurons can be estimated by fitting the model to a given set of spike data, such that each connection parameter represents the degree to which the firing activity of neurons is synchronized~\citep{Cocco2009, Posani2017}. A problem with this model is that the estimation results vary greatly depending on the time bin $\Delta$. To solve these problems, \cite{Terada2020} developed a kinetic Ising model that incorporates the effect of synaptic time delay and proposed a framework to optimize the bin width $\Delta$. 

\item {\bf Spiking neuron model:}
The spiking neuron model describes how a neuron generates spikes in response to external input~\citep{Kobayashi2009b, Gerstner2014}. 
It has also been proposed to estimate the connections between neurons by fitting a network of spiking neuron models to the parallel spike data~\citep{Cocco2009, Isomura2015, Ladenbauer2019}. It has also been proposed to estimate connections by training a spiking neural network with spike time-dependent plasticity (STDP) using measured spiking data~\citep{Moon2021}. 

\end{itemize}
\subsection{Available source codes}

As explained above, several techniques have been developed for estimating connections between neurons from spike data, and many of these techniques have been validated by applying them to simulated and experimental data. 
However, it often takes a lot of effort to implement these techniques programmatically. 
Therefore, it is convenient to use open source code. We list codes that are currently available (March 2024) in Table \ref{tab:codes}. 
Python and Matlab codes are becoming more popular. GLMCC and CoNNECT also have web applications that allow users to try analyzing spike data without programming. While TSPE, GLMCC, ExGLM, and CoNNECT are cross-correlation methods (Section 3.1.1), transfer entropy and CoTETE are based on information-theoretic measures (Section 3.1.2). GLMCC and CoNNECT were validated by applying them to simulated data from the spiking neuron model and the Hodgkin-Huxley model.  

%
%
\section{Future challenges}

We summarize the difficulties in estimating synaptic connections from spiking data and discuss possible future directions.

\begin{itemize}

\item {\bf False positives:} 
While reasonable estimation methods provide many true positives for synaptic connections, there are cases where they provide false positives. The false positives can be caused by functional similarity or common inputs. 
The false estimation due to functional similarity can be eliminated by using the shift predictor (Fig.\ref{fig:CC}b: \cite{Perkel1967}), and the false estimation due to common inputs can be eliminated by the GLMCC (Fig.\ref{fig:CC}e: \cite{Kobayashi2019}) or the GLM~\citep{Kulkarni2007}, but these methods are not perfect. Improving detection capability will continue to be a challenge. 

\item {\bf Competition of estimation methods:} 
We have seen that different approaches for estimating synaptic connections have been proposed and validated using simulated data.  It should be noted that the effective method may also depend on the simulation data.  Since model developers tend to prefer data that have yielded favorable results, it is desirable that a third party organize a competition to estimate synaptic connections from multicellular spike trains using experimental data in which the true connectivity is available for validation.

\item {\bf Network analysis:} 
Some studies have applied network analysis~\citep{Barabasi2016, Newman2018} to the estimated networks of neurons. \cite{Song2005} performed intracellular recordings from layer 5 pyramidal neurons in rat visual cortex and showed that the  connections between neurons have non-random features, such that bidirectional connectivity and three-neuron connectivity patterns are observed  more frequently than the random network (random graph). Furthermore, the strength of the connections follows a log-normal distribution, i.e. the synaptic connections consist of a few strong connections and many weak connections.  

\cite{Shimono2015} applied the transfer entropy method to spiking data measured with a 512-electrode array from cultured cells. They applied the network motif analysis~\citep{Milo2002} and the community detection algorithm~\citep{Blondel2008} to the inferred  networks. The results showed that the connections are more tightly connected than in the random network, there are hub neurons connected to many cells, and large clusters (communities) connected by two or three synapses are found. The functional significance of the network structures in information processing needs to be investigated in the future. 

\item {\bf Ca imaging data:} 
Ca imaging is a popular tool for measuring neural activity~\citep{Grienberger2012}, mainly because the number of neurons that can be recorded is huge; recently, neural activity from  16,000 neurons was recorded from multiple areas of awake mice, including sensory-motor areas~\citep{Ota2021}. 

However, the temporal resolution for spike time detection is still longer than the typical synaptic delay of a few milliseconds. Therefore, it is difficult to estimate synaptic connections from Ca imaging data. Nevertheless, several methods have been proposed to estimate synaptic connections directly from imaging data using GLM~\citep{Mishchencko2011} or transfer entropy~\citep{Stetter2012}. 
Furthermore, the first Neural Connectomics Challenge was organized with the goal of reconstructing monosynaptic connectivity from simulated Ca fluorescence signals~\citep{Guyon2014,de2018connectivity}. It would be essential to organize a similar challenge using experimental data. 

Since the activity of glial cells can be estimated from Ca imaging data, their influence on neurons can be modeled~\citep{Nakae2014}. Such techniques may allow a more accurate estimation of synaptic connections between neurons by removing the influence of glial cells. 

\item {\bf Improving the decoding accuracy:}  
It has been reported that information about interneuronal connectivity can be used to improve decoding performance. For example, the inferred connections can improve the reconstruction of visual information from retinal ganglion cell activity~\citep{Pillow2008} or the reconstruction of location information from hippocampal CA1 neuron activity~\citep{Posani2017}. Evaluating decoding performance may be useful to investigate the functional significance of synaptic connectivity.

\end{itemize}

\section{Conclusion}
In this paper we have reviewed the methods for inferring or estimating monosynaptic connectivity from parallel spike trains. 
First, we have clarified a variety of meanings of ``connectivity'' or ``connection'' as the term is used differently in different areas of neuroscience research, such as structural connectivity, monosynaptic connectivity, and functional connectivity. Structural connectivity is defined as the anatomical connectivity between neurons or between brain regions. Anatomical connection is a sufficient condition, but it does not necessarily guarantee the presence of a monosynaptic connection. In contrast, functional connectivity simply means that there is a significant correlation of brain activity, which does not necessarily mean that there are synaptic connections between them. 

Next, we focused on studies that infer monosynaptic connections from parallel spike trains of multiple neurons. 
As a result of efforts to refine the method, the accuracy of identifying monosynaptic connections seems to have improved dramatically in recent years. Nevertheless, it should be noted that the inference of synaptic connections will never be perfect~\citep{Brody1999,Gerstein1989,Stevenson2008a} or rule out the possibility of a connection based on the absence of a correlation~\citep{Stevenson2023}. To confirm the existence of a monosynaptic connection, one must either stimulate the presynaptic neuron by juxtacellular/micro-stimulation or optogenetic stimulation~\citep{Swadlow1995, English2017}, or manipulate the synaptic connections using synaptic blockers. However, due to the high cost of this experiment, it is impractical to do this for all neuron pairs. Therefore, a technique to estimate synaptic connections from spiking data would be useful to narrow down the number of neuron pairs to test. 

%
%
%
%
\section*{Acknowledgments} 
This research was motivated by a special session at the annual Japanese neuroscience meeting held in Sendai, Japan on August 2023. 
We thank the speakers and participants of the special session and the financial support from 2023 JNS NSR/Elsevier Sponsored Symposium. 
This research was partially supported by the Japan Society for the Promotion of Science (JSPS) KAKENHI (Nos. JP18K11560, JP19H01133, JP21H03559, JP21H04571, and JP22H03695) and AMED (No. JP21wm0525004 and JP223fa627001) to R.K, and JSPS KAKENHI No. 22H05163 to S.S. 
%
%



\bibliographystyle{elsarticle-harv} 
\bibliography{refs_GLMCC}

\begin{thebibliography}{110}
\expandafter\ifx\csname natexlab\endcsname\relax\def\natexlab#1{#1}\fi
\providecommand{\url}[1]{\texttt{#1}}
\providecommand{\href}[2]{#2}
\providecommand{\path}[1]{#1}
\providecommand{\DOIprefix}{doi:}
\providecommand{\ArXivprefix}{arXiv:}
\providecommand{\URLprefix}{URL: }
\providecommand{\Pubmedprefix}{pmid:}
\providecommand{\doi}[1]{\href{http://dx.doi.org/#1}{\path{#1}}}
\providecommand{\Pubmed}[1]{\href{pmid:#1}{\path{#1}}}
\providecommand{\bibinfo}[2]{#2}
\ifx\xfnm\relax \def\xfnm[#1]{\unskip,\space#1}\fi
\bibitem[{de~Abril et~al.(2018)de~Abril, Yoshimoto and
  Doya}]{de2018connectivity}
\bibinfo{author}{de~Abril, I.M.}, \bibinfo{author}{Yoshimoto, J.},
  \bibinfo{author}{Doya, K.}, \bibinfo{year}{2018}.
\newblock \bibinfo{title}{Connectivity inference from neural recording data:
  Challenges, mathematical bases and research directions}.
\newblock \bibinfo{journal}{Neural Networks} \bibinfo{volume}{102},
  \bibinfo{pages}{120--137}.
\bibitem[{Aertsen et~al.(1989)Aertsen, Gerstein, Habib and Palm}]{Aertsen1989}
\bibinfo{author}{Aertsen, A.}, \bibinfo{author}{Gerstein, G.},
  \bibinfo{author}{Habib, M.}, \bibinfo{author}{Palm, G.},
  \bibinfo{year}{1989}.
\newblock \bibinfo{title}{Dynamics of neuronal firing correlation: modulation
  of" effective connectivity"}.
\newblock \bibinfo{journal}{J. Neurophysiol.} \bibinfo{volume}{61},
  \bibinfo{pages}{900--917}.
\bibitem[{Amarasingham et~al.(2012)Amarasingham, Harrison, Hatsopoulos and
  Geman}]{Amarasingham2012}
\bibinfo{author}{Amarasingham, A.}, \bibinfo{author}{Harrison, M.T.},
  \bibinfo{author}{Hatsopoulos, N.G.}, \bibinfo{author}{Geman, S.},
  \bibinfo{year}{2012}.
\newblock \bibinfo{title}{Conditional modeling and the jitter method of spike
  resampling}.
\newblock \bibinfo{journal}{J. Neurophysiol.} \bibinfo{volume}{107},
  \bibinfo{pages}{517--531}.
\bibitem[{Antonello et~al.(2022)Antonello, Varley, Beggs, Porcionatto, Sporns
  and Faber}]{Antonello2022}
\bibinfo{author}{Antonello, P.C.}, \bibinfo{author}{Varley, T.F.},
  \bibinfo{author}{Beggs, J.}, \bibinfo{author}{Porcionatto, M.},
  \bibinfo{author}{Sporns, O.}, \bibinfo{author}{Faber, J.},
  \bibinfo{year}{2022}.
\newblock \bibinfo{title}{Self-organization of in vitro neuronal assemblies
  drives to complex network topology}.
\newblock \bibinfo{journal}{Elife} \bibinfo{volume}{11},
  \bibinfo{pages}{e74921}.
\bibitem[{Barabasi(2018)}]{Barabasi2016}
\bibinfo{author}{Barabasi, A.L.}, \bibinfo{year}{2018}.
\newblock \bibinfo{title}{Network Science}.
\newblock \bibinfo{publisher}{Oxford university press}.
\bibitem[{Barth\'o et~al.(2004)Barth\'o, Hirase, Monconduit, Zugaro, Harris and
  Buzs\'aki}]{Bartho2004}
\bibinfo{author}{Barth\'o, P.}, \bibinfo{author}{Hirase, H.},
  \bibinfo{author}{Monconduit, L.}, \bibinfo{author}{Zugaro, M.},
  \bibinfo{author}{Harris, K.D.}, \bibinfo{author}{Buzs\'aki, G.},
  \bibinfo{year}{2004}.
\newblock \bibinfo{title}{Characterization of neocortical principal cells and
  interneurons by network interactions and extracellular features}.
\newblock \bibinfo{journal}{J. Neurophysiol.} \bibinfo{volume}{92},
  \bibinfo{pages}{600--608}.
\bibitem[{Berry and Meister(1998)}]{Berry1998}
\bibinfo{author}{Berry, M.J.}, \bibinfo{author}{Meister, M.},
  \bibinfo{year}{1998}.
\newblock \bibinfo{title}{Refractoriness and neural precision}.
\newblock \bibinfo{journal}{J. Neurosci.} \bibinfo{volume}{18},
  \bibinfo{pages}{2200--2211}.
\bibitem[{Binzegger et~al.(2004)Binzegger, Douglas and Martin}]{Binzegger2004}
\bibinfo{author}{Binzegger, T.}, \bibinfo{author}{Douglas, R.J.},
  \bibinfo{author}{Martin, K.A.}, \bibinfo{year}{2004}.
\newblock \bibinfo{title}{A quantitative map of the circuit of cat primary
  visual cortex}.
\newblock \bibinfo{journal}{J. Neurosci.} \bibinfo{volume}{24},
  \bibinfo{pages}{8441--8453}.
\bibitem[{Blondel et~al.(2008)Blondel, Guillaume, Lambiotte and
  Lefebvre}]{Blondel2008}
\bibinfo{author}{Blondel, V.D.}, \bibinfo{author}{Guillaume, J.L.},
  \bibinfo{author}{Lambiotte, R.}, \bibinfo{author}{Lefebvre, E.},
  \bibinfo{year}{2008}.
\newblock \bibinfo{title}{Fast unfolding of communities in large networks}.
\newblock \bibinfo{journal}{J. Stat. Mech. Theor. Exp.} \bibinfo{volume}{2008},
  \bibinfo{pages}{P10008}.
\bibitem[{Brillinger(1975)}]{Brillinger1975}
\bibinfo{author}{Brillinger, D.R.}, \bibinfo{year}{1975}.
\newblock \bibinfo{title}{The identification of point process systems}.
\newblock \bibinfo{journal}{The Annals of Probability} ,
  \bibinfo{pages}{909--924}.
\bibitem[{Brillinger et~al.(1976)Brillinger, Bryant~Jr and
  Segundo}]{Brillinger1976}
\bibinfo{author}{Brillinger, D.R.}, \bibinfo{author}{Bryant~Jr, H.L.},
  \bibinfo{author}{Segundo, J.P.}, \bibinfo{year}{1976}.
\newblock \bibinfo{title}{Identification of synaptic interactions}.
\newblock \bibinfo{journal}{Biol. Cybern.} \bibinfo{volume}{22},
  \bibinfo{pages}{213--228}.
\bibitem[{Brody(1999)}]{Brody1999}
\bibinfo{author}{Brody, C.D.}, \bibinfo{year}{1999}.
\newblock \bibinfo{title}{Correlations without synchrony}.
\newblock \bibinfo{journal}{Neural computation} \bibinfo{volume}{11},
  \bibinfo{pages}{1537--1551}.
\bibitem[{Cai et~al.(2017)Cai, Neveu, Baxter, Byrne and Aazhang}]{Cai2017}
\bibinfo{author}{Cai, Z.}, \bibinfo{author}{Neveu, C.L.},
  \bibinfo{author}{Baxter, D.A.}, \bibinfo{author}{Byrne, J.H.},
  \bibinfo{author}{Aazhang, B.}, \bibinfo{year}{2017}.
\newblock \bibinfo{title}{Inferring neuronal network functional connectivity
  with directed information}.
\newblock \bibinfo{journal}{J. Neurophysiol.} \bibinfo{volume}{118},
  \bibinfo{pages}{1055--1069}.
\bibitem[{Chen and Brown(2022)}]{Chen2022}
\bibinfo{author}{Chen, Z.}, \bibinfo{author}{Brown, E.N.},
  \bibinfo{year}{2022}.
\newblock \bibinfo{title}{Generalized linear models for point process analyses
  of neural spiking activity}, in: \bibinfo{booktitle}{Encyclopedia of
  Computational Neuroscience}. \bibinfo{publisher}{Springer}, pp.
  \bibinfo{pages}{1510--1513}.
\bibitem[{Chen et~al.(2011)Chen, Putrino, Ghosh, Barbieri and Brown}]{Chen2011}
\bibinfo{author}{Chen, Z.}, \bibinfo{author}{Putrino, D.F.},
  \bibinfo{author}{Ghosh, S.}, \bibinfo{author}{Barbieri, R.},
  \bibinfo{author}{Brown, E.N.}, \bibinfo{year}{2011}.
\newblock \bibinfo{title}{Statistical inference for assessing functional
  connectivity of neuronal ensembles with sparse spiking data}.
\newblock \bibinfo{journal}{IEEE Trans. Neural Syst. Rehabil. Eng.}
  \bibinfo{volume}{19}, \bibinfo{pages}{121--135}.
\bibitem[{Chornoboy et~al.(1988)Chornoboy, Schramm and Karr}]{Chornoboy1988}
\bibinfo{author}{Chornoboy, E.}, \bibinfo{author}{Schramm, L.},
  \bibinfo{author}{Karr, A.}, \bibinfo{year}{1988}.
\newblock \bibinfo{title}{Maximum likelihood identification of neural point
  process systems}.
\newblock \bibinfo{journal}{Biol. Cybern.} \bibinfo{volume}{59},
  \bibinfo{pages}{265--275}.
\bibitem[{Cocco et~al.(2009)Cocco, Leibler and Monasson}]{Cocco2009}
\bibinfo{author}{Cocco, S.}, \bibinfo{author}{Leibler, S.},
  \bibinfo{author}{Monasson, R.}, \bibinfo{year}{2009}.
\newblock \bibinfo{title}{Neuronal couplings between retinal ganglion cells
  inferred by efficient inverse statistical physics methods}.
\newblock \bibinfo{journal}{Proc. Natl. Acad. Sci.} \bibinfo{volume}{106},
  \bibinfo{pages}{14058--14062}.
\bibitem[{Cover and Thomas(1991)}]{Cover1991}
\bibinfo{author}{Cover, M.}, \bibinfo{author}{Thomas, J.},
  \bibinfo{year}{1991}.
\newblock \bibinfo{title}{Elements of information theory}.
\newblock \bibinfo{publisher}{John Wiley \& Sons}.
\bibitem[{De~Blasi et~al.(2019)De~Blasi, Ciba, Bahmer and Thielemann}]{De2019}
\bibinfo{author}{De~Blasi, S.}, \bibinfo{author}{Ciba, M.},
  \bibinfo{author}{Bahmer, A.}, \bibinfo{author}{Thielemann, C.},
  \bibinfo{year}{2019}.
\newblock \bibinfo{title}{Total spiking probability edges: A cross-correlation
  based method for effective connectivity estimation of cortical spiking
  neurons}.
\newblock \bibinfo{journal}{J. Neurosci. Methods} \bibinfo{volume}{312},
  \bibinfo{pages}{169--181}.
\bibitem[{Endo et~al.(2021)Endo, Kobayashi, Bartolo, Averbeck, Sugase-Miyamoto,
  Hayashi, Kawano, Richmond and Shinomoto}]{Endo2021}
\bibinfo{author}{Endo, D.}, \bibinfo{author}{Kobayashi, R.},
  \bibinfo{author}{Bartolo, R.}, \bibinfo{author}{Averbeck, B.B.},
  \bibinfo{author}{Sugase-Miyamoto, Y.}, \bibinfo{author}{Hayashi, K.},
  \bibinfo{author}{Kawano, K.}, \bibinfo{author}{Richmond, B.J.},
  \bibinfo{author}{Shinomoto, S.}, \bibinfo{year}{2021}.
\newblock \bibinfo{title}{A convolutional neural network for estimating
  synaptic connectivity from spike trains}.
\newblock \bibinfo{journal}{Sci. Rep.} \bibinfo{volume}{11},
  \bibinfo{pages}{12087}.
\bibitem[{English et~al.(2017)English, McKenzie, Evans, Kim, Yoon and
  Buzs{\'a}ki}]{English2017}
\bibinfo{author}{English, D.F.}, \bibinfo{author}{McKenzie, S.},
  \bibinfo{author}{Evans, T.}, \bibinfo{author}{Kim, K.},
  \bibinfo{author}{Yoon, E.}, \bibinfo{author}{Buzs{\'a}ki, G.},
  \bibinfo{year}{2017}.
\newblock \bibinfo{title}{Pyramidal cell-interneuron circuit architecture and
  dynamics in hippocampal networks}.
\newblock \bibinfo{journal}{Neuron} \bibinfo{volume}{96},
  \bibinfo{pages}{505--520}.
\bibitem[{Ermentrout(1998)}]{Ermentrout1998}
\bibinfo{author}{Ermentrout, B.}, \bibinfo{year}{1998}.
\newblock \bibinfo{title}{Linearization of fi curves by adaptation}.
\newblock \bibinfo{journal}{Neural Comput.} \bibinfo{volume}{10},
  \bibinfo{pages}{1721--1729}.
\bibitem[{Evarts(1968)}]{evarts1968relation}
\bibinfo{author}{Evarts, E.V.}, \bibinfo{year}{1968}.
\newblock \bibinfo{title}{Relation of pyramidal tract activity to force exerted
  during voluntary movement.}
\newblock \bibinfo{journal}{Journal of neurophysiology} \bibinfo{volume}{31},
  \bibinfo{pages}{14--27}.
\bibitem[{Friston(1994)}]{Friston1994}
\bibinfo{author}{Friston, K.J.}, \bibinfo{year}{1994}.
\newblock \bibinfo{title}{Functional and effective connectivity in
  neuroimaging: a synthesis}.
\newblock \bibinfo{journal}{Hum. Brain Mapp.} \bibinfo{volume}{2},
  \bibinfo{pages}{56--78}.
\bibitem[{Fujisawa et~al.(2008)Fujisawa, Amarasingham, Harrison and
  Buzs{\'a}ki}]{Fujisawa2008}
\bibinfo{author}{Fujisawa, S.}, \bibinfo{author}{Amarasingham, A.},
  \bibinfo{author}{Harrison, M.T.}, \bibinfo{author}{Buzs{\'a}ki, G.},
  \bibinfo{year}{2008}.
\newblock \bibinfo{title}{Behavior-dependent short-term assembly dynamics in
  the medial prefrontal cortex}.
\newblock \bibinfo{journal}{Nat. Neurosci.} \bibinfo{volume}{11},
  \bibinfo{pages}{823}.
\bibitem[{Fujita et~al.(1992)Fujita, Tanaka, Ito and Cheng}]{fujita1992columns}
\bibinfo{author}{Fujita, I.}, \bibinfo{author}{Tanaka, K.},
  \bibinfo{author}{Ito, M.}, \bibinfo{author}{Cheng, K.}, \bibinfo{year}{1992}.
\newblock \bibinfo{title}{Columns for visual features of objects in monkey
  inferotemporal cortex}.
\newblock \bibinfo{journal}{Nature} \bibinfo{volume}{360},
  \bibinfo{pages}{343--346}.
\bibitem[{Funahashi and Inoue(2000)}]{Funahashi2000}
\bibinfo{author}{Funahashi, S.}, \bibinfo{author}{Inoue, M.},
  \bibinfo{year}{2000}.
\newblock \bibinfo{title}{Neuronal interactions related to working memory
  processes in the primate prefrontal cortex revealed by cross-correlation
  analysis}.
\newblock \bibinfo{journal}{Cerebral Cortex} \bibinfo{volume}{10},
  \bibinfo{pages}{535--551}.
\bibitem[{Garofalo et~al.(2009)Garofalo, Nieus, Massobrio and
  Martinoia}]{Garofalo2009}
\bibinfo{author}{Garofalo, M.}, \bibinfo{author}{Nieus, T.},
  \bibinfo{author}{Massobrio, P.}, \bibinfo{author}{Martinoia, S.},
  \bibinfo{year}{2009}.
\newblock \bibinfo{title}{Evaluation of the performance of information
  theory-based methods and cross-correlation to estimate the functional
  connectivity in cortical networks}.
\newblock \bibinfo{journal}{PloS ONE} \bibinfo{volume}{4},
  \bibinfo{pages}{e6482}.
\bibitem[{Gerhard et~al.(2013)Gerhard, Kispersky, Gutierrez, Marder, Kramer and
  Eden}]{Gerhard2013}
\bibinfo{author}{Gerhard, F.}, \bibinfo{author}{Kispersky, T.},
  \bibinfo{author}{Gutierrez, G.J.}, \bibinfo{author}{Marder, E.},
  \bibinfo{author}{Kramer, M.}, \bibinfo{author}{Eden, U.},
  \bibinfo{year}{2013}.
\newblock \bibinfo{title}{Successful reconstruction of a physiological circuit
  with known connectivity from spiking activity alone}.
\newblock \bibinfo{journal}{PLoS Comput. Biol.} \bibinfo{volume}{9},
  \bibinfo{pages}{e1003138}.
\bibitem[{Gerstein et~al.(1989)Gerstein, Bedenbaugh and Aertsen}]{Gerstein1989}
\bibinfo{author}{Gerstein, G.L.}, \bibinfo{author}{Bedenbaugh, P.},
  \bibinfo{author}{Aertsen, A.M.}, \bibinfo{year}{1989}.
\newblock \bibinfo{title}{Neuronal assemblies}.
\newblock \bibinfo{journal}{IEEE Transactions on Biomedical Engineering}
  \bibinfo{volume}{36}, \bibinfo{pages}{4--14}.
\bibitem[{Gerstein and Perkel(1969)}]{Gerstein1969}
\bibinfo{author}{Gerstein, G.L.}, \bibinfo{author}{Perkel, D.H.},
  \bibinfo{year}{1969}.
\newblock \bibinfo{title}{Simultaneously recorded trains of action potentials:
  analysis and functional interpretation}.
\newblock \bibinfo{journal}{Science} \bibinfo{volume}{164},
  \bibinfo{pages}{828--830}.
\bibitem[{Gerstner et~al.(2014)Gerstner, Kistler, Naud and
  Paninski}]{Gerstner2014}
\bibinfo{author}{Gerstner, W.}, \bibinfo{author}{Kistler, W.M.},
  \bibinfo{author}{Naud, R.}, \bibinfo{author}{Paninski, L.},
  \bibinfo{year}{2014}.
\newblock \bibinfo{title}{Neuronal dynamics: From single neurons to networks
  and models of cognition}.
\newblock \bibinfo{publisher}{Cambridge University Press}.
\bibitem[{Gerwinn et~al.(2010)Gerwinn, Macke and Bethge}]{Gerwinn2010}
\bibinfo{author}{Gerwinn, S.}, \bibinfo{author}{Macke, J.H.},
  \bibinfo{author}{Bethge, M.}, \bibinfo{year}{2010}.
\newblock \bibinfo{title}{Bayesian inference for generalized linear models for
  spiking neurons}.
\newblock \bibinfo{journal}{Front. Comput. Neurosci.} \bibinfo{volume}{4},
  \bibinfo{pages}{1299}.
\bibitem[{Granger(1969)}]{Granger1969}
\bibinfo{author}{Granger, C.W.}, \bibinfo{year}{1969}.
\newblock \bibinfo{title}{Investigating causal relations by econometric models
  and cross-spectral methods}.
\newblock \bibinfo{journal}{Econometrica} , \bibinfo{pages}{424--438}.
\bibitem[{Grienberger and Konnerth(2012)}]{Grienberger2012}
\bibinfo{author}{Grienberger, C.}, \bibinfo{author}{Konnerth, A.},
  \bibinfo{year}{2012}.
\newblock \bibinfo{title}{Imaging calcium in neurons}.
\newblock \bibinfo{journal}{Neuron} \bibinfo{volume}{73},
  \bibinfo{pages}{862--885}.
\bibitem[{Guyon et~al.(2014)Guyon, Battaglia, Guyon, Lemaire, Orlandi, Ray,
  Saeed, Soriano, Statnikov and Stetter}]{Guyon2014}
\bibinfo{author}{Guyon, I.}, \bibinfo{author}{Battaglia, D.},
  \bibinfo{author}{Guyon, A.}, \bibinfo{author}{Lemaire, V.},
  \bibinfo{author}{Orlandi, J.G.}, \bibinfo{author}{Ray, B.},
  \bibinfo{author}{Saeed, M.}, \bibinfo{author}{Soriano, J.},
  \bibinfo{author}{Statnikov, A.}, \bibinfo{author}{Stetter, O.},
  \bibinfo{year}{2014}.
\newblock \bibinfo{title}{Design of the first neuronal connectomics challenge:
  From imaging to connectivity}, in: \bibinfo{booktitle}{2014 International
  Joint Conference on Neural Networks (IJCNN)}, \bibinfo{organization}{IEEE}.
  pp. \bibinfo{pages}{2600--2607}.
\bibitem[{Hawkes(1971)}]{Hawkes1971}
\bibinfo{author}{Hawkes, A.G.}, \bibinfo{year}{1971}.
\newblock \bibinfo{title}{Spectra of some self-exciting and mutually exciting
  point processes}.
\newblock \bibinfo{journal}{Biometrika} \bibinfo{volume}{58},
  \bibinfo{pages}{83--90}.
\bibitem[{Helmstaedter et~al.(2013)Helmstaedter, Briggman, Turaga, Jain, Seung
  and Denk}]{Helmstaedter2013}
\bibinfo{author}{Helmstaedter, M.}, \bibinfo{author}{Briggman, K.L.},
  \bibinfo{author}{Turaga, S.C.}, \bibinfo{author}{Jain, V.},
  \bibinfo{author}{Seung, H.S.}, \bibinfo{author}{Denk, W.},
  \bibinfo{year}{2013}.
\newblock \bibinfo{title}{Connectomic reconstruction of the inner plexiform
  layer in the mouse retina}.
\newblock \bibinfo{journal}{Nature} \bibinfo{volume}{500},
  \bibinfo{pages}{168--174}.
\bibitem[{van~den Heuvel and Sporns(2019)}]{Van2019}
\bibinfo{author}{van~den Heuvel, M.P.}, \bibinfo{author}{Sporns, O.},
  \bibinfo{year}{2019}.
\newblock \bibinfo{title}{A cross-disorder connectome landscape of brain
  dysconnectivity}.
\newblock \bibinfo{journal}{Nat. Rev. Neurosci.} \bibinfo{volume}{20},
  \bibinfo{pages}{435--446}.
\bibitem[{Hubel and Wiesel(1977)}]{hubel1977ferrier}
\bibinfo{author}{Hubel, D.H.}, \bibinfo{author}{Wiesel, T.N.},
  \bibinfo{year}{1977}.
\newblock \bibinfo{title}{Ferrier lecture-functional architecture of macaque
  monkey visual cortex}.
\newblock \bibinfo{journal}{Proceedings of the Royal Society of London. Series
  B. Biological Sciences} \bibinfo{volume}{198}, \bibinfo{pages}{1--59}.
\bibitem[{Isomura et~al.(2015)Isomura, Ogawa, Kotani and Jimbo}]{Isomura2015}
\bibinfo{author}{Isomura, T.}, \bibinfo{author}{Ogawa, Y.},
  \bibinfo{author}{Kotani, K.}, \bibinfo{author}{Jimbo, Y.},
  \bibinfo{year}{2015}.
\newblock \bibinfo{title}{Accurate connection strength estimation based on
  variational bayes for detecting synaptic plasticity}.
\newblock \bibinfo{journal}{Neural Comput.} \bibinfo{volume}{27},
  \bibinfo{pages}{819--844}.
\bibitem[{Ito et~al.(2011)Ito, Hansen, Heiland, Lumsdaine, Litke and
  Beggs}]{Ito2011}
\bibinfo{author}{Ito, S.}, \bibinfo{author}{Hansen, M.E.},
  \bibinfo{author}{Heiland, R.}, \bibinfo{author}{Lumsdaine, A.},
  \bibinfo{author}{Litke, A.M.}, \bibinfo{author}{Beggs, J.M.},
  \bibinfo{year}{2011}.
\newblock \bibinfo{title}{Extending transfer entropy improves identification of
  effective connectivity in a spiking cortical network model}.
\newblock \bibinfo{journal}{PloS ONE} \bibinfo{volume}{6},
  \bibinfo{pages}{e27431}.
\bibitem[{Izhikevich and Edelman(2008)}]{Izhikevich2008}
\bibinfo{author}{Izhikevich, E.M.}, \bibinfo{author}{Edelman, G.M.},
  \bibinfo{year}{2008}.
\newblock \bibinfo{title}{Large-scale model of mammalian thalamocortical
  systems}.
\newblock \bibinfo{journal}{Proc. Natl. Acad. Sci. USA} \bibinfo{volume}{105},
  \bibinfo{pages}{3593--3598}.
\bibitem[{Jolivet et~al.(2006)Jolivet, Rauch, L{\"u}scher and
  Gerstner}]{Jolivet2006}
\bibinfo{author}{Jolivet, R.}, \bibinfo{author}{Rauch, A.},
  \bibinfo{author}{L{\"u}scher, H.R.}, \bibinfo{author}{Gerstner, W.},
  \bibinfo{year}{2006}.
\newblock \bibinfo{title}{Predicting spike timing of neocortical pyramidal
  neurons by simple threshold models}.
\newblock \bibinfo{journal}{J. Comput. Neurosci.} \bibinfo{volume}{21},
  \bibinfo{pages}{35--49}.
\bibitem[{Kim et~al.(2011)Kim, Putrino, Ghosh and Brown}]{Kim2011}
\bibinfo{author}{Kim, S.}, \bibinfo{author}{Putrino, D.},
  \bibinfo{author}{Ghosh, S.}, \bibinfo{author}{Brown, E.N.},
  \bibinfo{year}{2011}.
\newblock \bibinfo{title}{A granger causality measure for point process models
  of ensemble neural spiking activity}.
\newblock \bibinfo{journal}{PLoS Comput. Biol.} \bibinfo{volume}{7},
  \bibinfo{pages}{e1001110}.
\bibitem[{Kobayashi(2009)}]{Kobayashi2009a}
\bibinfo{author}{Kobayashi, R.}, \bibinfo{year}{2009}.
\newblock \bibinfo{title}{The influence of firing mechanisms on gain
  modulation}.
\newblock \bibinfo{journal}{J. Stat. Mech. Theor. Exp.} \bibinfo{volume}{2009},
  \bibinfo{pages}{P01017}.
\bibitem[{Kobayashi and Kitano(2013)}]{Kobayashi2013}
\bibinfo{author}{Kobayashi, R.}, \bibinfo{author}{Kitano, K.},
  \bibinfo{year}{2013}.
\newblock \bibinfo{title}{Impact of network topology on inference of synaptic
  connectivity from multi-neuronal spike data simulated by a large-scale
  cortical network model}.
\newblock \bibinfo{journal}{J. Comput. Neurosci.} \bibinfo{volume}{35},
  \bibinfo{pages}{109--124}.
\bibitem[{Kobayashi and Kitano(2016)}]{Kobayashi2016}
\bibinfo{author}{Kobayashi, R.}, \bibinfo{author}{Kitano, K.},
  \bibinfo{year}{2016}.
\newblock \bibinfo{title}{Impact of slow {K}$^+$ currents on spike generation
  can be described by an adaptive threshold model}.
\newblock \bibinfo{journal}{J. Comput. Neurosci.} \bibinfo{volume}{40},
  \bibinfo{pages}{347--362}.
\bibitem[{Kobayashi et~al.(2019)Kobayashi, Kurita, Kurth, Kitano, Mizuseki,
  Diesmann, Richmond and Shinomoto}]{Kobayashi2019}
\bibinfo{author}{Kobayashi, R.}, \bibinfo{author}{Kurita, S.},
  \bibinfo{author}{Kurth, A.}, \bibinfo{author}{Kitano, K.},
  \bibinfo{author}{Mizuseki, K.}, \bibinfo{author}{Diesmann, M.},
  \bibinfo{author}{Richmond, B.J.}, \bibinfo{author}{Shinomoto, S.},
  \bibinfo{year}{2019}.
\newblock \bibinfo{title}{Reconstructing neuronal circuitry from parallel spike
  trains}.
\newblock \bibinfo{journal}{Nat. Commun.} \bibinfo{volume}{10},
  \bibinfo{pages}{4468}.
\bibitem[{Kobayashi and Shinomoto(2007)}]{Kobayashi2007}
\bibinfo{author}{Kobayashi, R.}, \bibinfo{author}{Shinomoto, S.},
  \bibinfo{year}{2007}.
\newblock \bibinfo{title}{State space method for predicting the spike times of
  a neuron}.
\newblock \bibinfo{journal}{Phys. Rev. E} \bibinfo{volume}{75},
  \bibinfo{pages}{011925}.
\bibitem[{Kobayashi et~al.(2009)Kobayashi, Tsubo and
  Shinomoto}]{Kobayashi2009b}
\bibinfo{author}{Kobayashi, R.}, \bibinfo{author}{Tsubo, Y.},
  \bibinfo{author}{Shinomoto, S.}, \bibinfo{year}{2009}.
\newblock \bibinfo{title}{Made-to-order spiking neuron model equipped with a
  multi-timescale adaptive threshold}.
\newblock \bibinfo{journal}{Front. Comput. Neurosci.} \bibinfo{volume}{3},
  \bibinfo{pages}{9}.
\bibitem[{Koch(2004)}]{Koch2004}
\bibinfo{author}{Koch, C.}, \bibinfo{year}{2004}.
\newblock \bibinfo{title}{Biophysics of computation: information processing in
  single neurons}.
\newblock \bibinfo{publisher}{Oxford university press}.
\bibitem[{Kulkarni and Paninski(2007)}]{Kulkarni2007}
\bibinfo{author}{Kulkarni, J.E.}, \bibinfo{author}{Paninski, L.},
  \bibinfo{year}{2007}.
\newblock \bibinfo{title}{Common-input models for multiple neural spike-train
  data}.
\newblock \bibinfo{journal}{Network} \bibinfo{volume}{18},
  \bibinfo{pages}{375--407}.
\bibitem[{Kuroda et~al.(2011)Kuroda, Ashizawa and Ikeguchi}]{Kuroda2011}
\bibinfo{author}{Kuroda, K.}, \bibinfo{author}{Ashizawa, T.},
  \bibinfo{author}{Ikeguchi, T.}, \bibinfo{year}{2011}.
\newblock \bibinfo{title}{Estimation of network structures only from spike
  sequences}.
\newblock \bibinfo{journal}{Physica A} \bibinfo{volume}{390},
  \bibinfo{pages}{4002--4011}.
\bibitem[{Ladenbauer et~al.(2019)Ladenbauer, McKenzie, English, Hagens and
  Ostojic}]{Ladenbauer2019}
\bibinfo{author}{Ladenbauer, J.}, \bibinfo{author}{McKenzie, S.},
  \bibinfo{author}{English, D.F.}, \bibinfo{author}{Hagens, O.},
  \bibinfo{author}{Ostojic, S.}, \bibinfo{year}{2019}.
\newblock \bibinfo{title}{Inferring and validating mechanistic models of neural
  microcircuits based on spike-train data}.
\newblock \bibinfo{journal}{Nat. Commun.} \bibinfo{volume}{10},
  \bibinfo{pages}{4933}.
\bibitem[{Lambert et~al.(2018)Lambert, Tuleau-Malot, Bessaih, Rivoirard,
  Bouret, Leresche and Reynaud-Bouret}]{Lambert2018}
\bibinfo{author}{Lambert, R.C.}, \bibinfo{author}{Tuleau-Malot, C.},
  \bibinfo{author}{Bessaih, T.}, \bibinfo{author}{Rivoirard, V.},
  \bibinfo{author}{Bouret, Y.}, \bibinfo{author}{Leresche, N.},
  \bibinfo{author}{Reynaud-Bouret, P.}, \bibinfo{year}{2018}.
\newblock \bibinfo{title}{Reconstructing the functional connectivity of
  multiple spike trains using hawkes models}.
\newblock \bibinfo{journal}{J. Neurosci. Methods} \bibinfo{volume}{297},
  \bibinfo{pages}{9--21}.
\bibitem[{Liew et~al.(2021)Liew, Pala, Whitmire, Stoy, Forest and
  Stanley}]{Liew2021}
\bibinfo{author}{Liew, Y.J.}, \bibinfo{author}{Pala, A.},
  \bibinfo{author}{Whitmire, C.J.}, \bibinfo{author}{Stoy, W.A.},
  \bibinfo{author}{Forest, C.R.}, \bibinfo{author}{Stanley, G.B.},
  \bibinfo{year}{2021}.
\newblock \bibinfo{title}{Inferring thalamocortical monosynaptic connectivity
  in vivo}.
\newblock \bibinfo{journal}{J. Neurophysiol.} \bibinfo{volume}{125},
  \bibinfo{pages}{2408--2431}.
\bibitem[{Liu et~al.(2017)Liu, Schreyer, Onken, Rozenblit, Khani,
  Krishnamoorthy, Panzeri and Gollisch}]{Liu2017}
\bibinfo{author}{Liu, J.K.}, \bibinfo{author}{Schreyer, H.M.},
  \bibinfo{author}{Onken, A.}, \bibinfo{author}{Rozenblit, F.},
  \bibinfo{author}{Khani, M.H.}, \bibinfo{author}{Krishnamoorthy, V.},
  \bibinfo{author}{Panzeri, S.}, \bibinfo{author}{Gollisch, T.},
  \bibinfo{year}{2017}.
\newblock \bibinfo{title}{Inference of neuronal functional circuitry with
  spike-triggered non-negative matrix factorization}.
\newblock \bibinfo{journal}{Nat. Commun.} \bibinfo{volume}{8},
  \bibinfo{pages}{149}.
\bibitem[{Lynall et~al.(2010)Lynall, Bassett, Kerwin, McKenna, Kitzbichler,
  Muller and Bullmore}]{Lynall2010}
\bibinfo{author}{Lynall, M.E.}, \bibinfo{author}{Bassett, D.S.},
  \bibinfo{author}{Kerwin, R.}, \bibinfo{author}{McKenna, P.J.},
  \bibinfo{author}{Kitzbichler, M.}, \bibinfo{author}{Muller, U.},
  \bibinfo{author}{Bullmore, E.}, \bibinfo{year}{2010}.
\newblock \bibinfo{title}{Functional connectivity and brain networks in
  schizophrenia}.
\newblock \bibinfo{journal}{J. Neurosci.} \bibinfo{volume}{30},
  \bibinfo{pages}{9477--9487}.
\bibitem[{Milo et~al.(2002)Milo, Shen-Orr, Itzkovitz, Kashtan, Chklovskii and
  Alon}]{Milo2002}
\bibinfo{author}{Milo, R.}, \bibinfo{author}{Shen-Orr, S.},
  \bibinfo{author}{Itzkovitz, S.}, \bibinfo{author}{Kashtan, N.},
  \bibinfo{author}{Chklovskii, D.}, \bibinfo{author}{Alon, U.},
  \bibinfo{year}{2002}.
\newblock \bibinfo{title}{Network motifs: simple building blocks of complex
  networks}.
\newblock \bibinfo{journal}{Science} \bibinfo{volume}{298},
  \bibinfo{pages}{824--827}.
\bibitem[{Mishchencko et~al.(2011)Mishchencko, Vogelstein and
  Paninski}]{Mishchencko2011}
\bibinfo{author}{Mishchencko, Y.}, \bibinfo{author}{Vogelstein, J.T.},
  \bibinfo{author}{Paninski, L.}, \bibinfo{year}{2011}.
\newblock \bibinfo{title}{A bayesian approach for inferring neuronal
  connectivity from calcium fluorescent imaging data}.
\newblock \bibinfo{journal}{Ann. Appl. Stat.} \bibinfo{volume}{5},
  \bibinfo{pages}{1229--1261}.
\bibitem[{Mizuseki et~al.(2009)Mizuseki, Sirota, Pastalkova and
  Buzs{\'a}ki}]{Mizuseki2009}
\bibinfo{author}{Mizuseki, K.}, \bibinfo{author}{Sirota, A.},
  \bibinfo{author}{Pastalkova, E.}, \bibinfo{author}{Buzs{\'a}ki, G.},
  \bibinfo{year}{2009}.
\newblock \bibinfo{title}{Theta oscillations provide temporal windows for local
  circuit computation in the entorhinal-hippocampal loop}.
\newblock \bibinfo{journal}{Neuron} \bibinfo{volume}{64},
  \bibinfo{pages}{267--280}.
\bibitem[{Mizuseki et~al.(2013)Mizuseki, Sirota, Pastalkova, Diba and
  Buzs{\'a}ki}]{Mizuseki2013}
\bibinfo{author}{Mizuseki, K.}, \bibinfo{author}{Sirota, A.},
  \bibinfo{author}{Pastalkova, E.}, \bibinfo{author}{Diba, K.},
  \bibinfo{author}{Buzs{\'a}ki, G.}, \bibinfo{year}{2013}.
\newblock \bibinfo{title}{Multiple single unit recordings from different rat
  hippocampal and entorhinal regions while the animals were performing multiple
  behavioral tasks}.
\newblock \bibinfo{journal}{CRCNS org} .
\bibitem[{Moon et~al.(2021)Moon, Wu, Zhu and Lu}]{Moon2021}
\bibinfo{author}{Moon, J.}, \bibinfo{author}{Wu, Y.}, \bibinfo{author}{Zhu,
  X.}, \bibinfo{author}{Lu, W.D.}, \bibinfo{year}{2021}.
\newblock \bibinfo{title}{Neural connectivity inference with spike-timing
  dependent plasticity network}.
\newblock \bibinfo{journal}{Sci. China Inf. Sci.} \bibinfo{volume}{64},
  \bibinfo{pages}{160405}.
\bibitem[{Moore et~al.(1970)Moore, Segundo, Perkel and Levitan}]{Moore1970}
\bibinfo{author}{Moore, G.P.}, \bibinfo{author}{Segundo, J.P.},
  \bibinfo{author}{Perkel, D.H.}, \bibinfo{author}{Levitan, H.},
  \bibinfo{year}{1970}.
\newblock \bibinfo{title}{Statistical signs of synaptic interaction in
  neurons}.
\newblock \bibinfo{journal}{Biophys J.} \bibinfo{volume}{10},
  \bibinfo{pages}{876--900}.
\bibitem[{Nakae et~al.(2014)Nakae, Ikegaya, Ishikawa, Oba, Urakubo, Koyama and
  Ishii}]{Nakae2014}
\bibinfo{author}{Nakae, K.}, \bibinfo{author}{Ikegaya, Y.},
  \bibinfo{author}{Ishikawa, T.}, \bibinfo{author}{Oba, S.},
  \bibinfo{author}{Urakubo, H.}, \bibinfo{author}{Koyama, M.},
  \bibinfo{author}{Ishii, S.}, \bibinfo{year}{2014}.
\newblock \bibinfo{title}{A statistical method of identifying interactions in
  neuron--glia systems based on functional multicell {Ca}$^{2+}$ imaging}.
\newblock \bibinfo{journal}{PLoS Comput. Biol.} \bibinfo{volume}{10},
  \bibinfo{pages}{e1003949}.
\bibitem[{Nedungadi et~al.(2009)Nedungadi, Rangarajan, Jain and
  Ding}]{Nedungadi2009}
\bibinfo{author}{Nedungadi, A.G.}, \bibinfo{author}{Rangarajan, G.},
  \bibinfo{author}{Jain, N.}, \bibinfo{author}{Ding, M.}, \bibinfo{year}{2009}.
\newblock \bibinfo{title}{Analyzing multiple spike trains with nonparametric
  granger causality}.
\newblock \bibinfo{journal}{J. Comput. Neurosci.} \bibinfo{volume}{27},
  \bibinfo{pages}{55--64}.
\bibitem[{Newman(2018)}]{Newman2018}
\bibinfo{author}{Newman, M.}, \bibinfo{year}{2018}.
\newblock \bibinfo{title}{Networks}.
\newblock \bibinfo{publisher}{Oxford university press}.
\bibitem[{Nowak et~al.(1999)Nowak, Munk, James, Girard and Bullier}]{Nowak1999}
\bibinfo{author}{Nowak, L.}, \bibinfo{author}{Munk, M.},
  \bibinfo{author}{James, A.}, \bibinfo{author}{Girard, P.},
  \bibinfo{author}{Bullier, J.}, \bibinfo{year}{1999}.
\newblock \bibinfo{title}{Cross-correlation study of the temporal interactions
  between areas {V}1 and {V}2 of the macaque monkey}.
\newblock \bibinfo{journal}{J. Neurophysiol.} \bibinfo{volume}{81},
  \bibinfo{pages}{1057--1074}.
\bibitem[{Okatan et~al.(2005)Okatan, Wilson and Brown}]{Okatan2005}
\bibinfo{author}{Okatan, M.}, \bibinfo{author}{Wilson, M.A.},
  \bibinfo{author}{Brown, E.N.}, \bibinfo{year}{2005}.
\newblock \bibinfo{title}{Analyzing functional connectivity using a network
  likelihood model of ensemble neural spiking activity}.
\newblock \bibinfo{journal}{Neural Comput.} \bibinfo{volume}{17},
  \bibinfo{pages}{1927--1961}.
\bibitem[{Ota et~al.(2021)Ota, Oisi, Suzuki, Ikeda, Ito, Ito, Uwamori,
  Kobayashi, Kobayashi, Odagawa et~al.}]{Ota2021}
\bibinfo{author}{Ota, K.}, \bibinfo{author}{Oisi, Y.}, \bibinfo{author}{Suzuki,
  T.}, \bibinfo{author}{Ikeda, M.}, \bibinfo{author}{Ito, Y.},
  \bibinfo{author}{Ito, T.}, \bibinfo{author}{Uwamori, H.},
  \bibinfo{author}{Kobayashi, K.}, \bibinfo{author}{Kobayashi, M.},
  \bibinfo{author}{Odagawa, M.}, et~al., \bibinfo{year}{2021}.
\newblock \bibinfo{title}{Fast, cell-resolution, contiguous-wide two-photon
  imaging to reveal functional network architectures across multi-modal
  cortical areas}.
\newblock \bibinfo{journal}{Neuron} \bibinfo{volume}{109},
  \bibinfo{pages}{1810--1824}.
\bibitem[{Palm et~al.(1988)Palm, Aertsen and Gerstein}]{Palm1988}
\bibinfo{author}{Palm, G.}, \bibinfo{author}{Aertsen, A.},
  \bibinfo{author}{Gerstein, G.}, \bibinfo{year}{1988}.
\newblock \bibinfo{title}{On the significance of correlations among neuronal
  spike trains}.
\newblock \bibinfo{journal}{Biol. Cybern.} \bibinfo{volume}{59},
  \bibinfo{pages}{1--11}.
\bibitem[{Paninski(2004)}]{Paninski2004}
\bibinfo{author}{Paninski, L.}, \bibinfo{year}{2004}.
\newblock \bibinfo{title}{Maximum likelihood estimation of cascade
  point-process neural encoding models}.
\newblock \bibinfo{journal}{Network} \bibinfo{volume}{15},
  \bibinfo{pages}{243}.
\bibitem[{Pastore et~al.(2018)Pastore, Massobrio, Godjoski and
  Martinoia}]{Pastore2018}
\bibinfo{author}{Pastore, V.P.}, \bibinfo{author}{Massobrio, P.},
  \bibinfo{author}{Godjoski, A.}, \bibinfo{author}{Martinoia, S.},
  \bibinfo{year}{2018}.
\newblock \bibinfo{title}{Identification of excitatory-inhibitory links and
  network topology in large-scale neuronal assemblies from multi-electrode
  recordings}.
\newblock \bibinfo{journal}{PLoS Comput. Biol.} \bibinfo{volume}{14},
  \bibinfo{pages}{e1006381}.
\bibitem[{Perkel et~al.(1967)Perkel, Gerstein and Moore}]{Perkel1967}
\bibinfo{author}{Perkel, D.H.}, \bibinfo{author}{Gerstein, G.L.},
  \bibinfo{author}{Moore, G.P.}, \bibinfo{year}{1967}.
\newblock \bibinfo{title}{Neuronal spike trains and stochastic point processes:
  {II}. simultaneous spike trains}.
\newblock \bibinfo{journal}{Biophys J.} \bibinfo{volume}{7},
  \bibinfo{pages}{419--440}.
\bibitem[{Pernice and Rotter(2013)}]{Pernice2013}
\bibinfo{author}{Pernice, V.}, \bibinfo{author}{Rotter, S.},
  \bibinfo{year}{2013}.
\newblock \bibinfo{title}{Reconstruction of sparse connectivity in neural
  networks from spike train covariances}.
\newblock \bibinfo{journal}{J. Stat. Mech. Theor. Exp.} ,
  \bibinfo{pages}{P03008}.
\bibitem[{Pillow et~al.(2008)Pillow, Shlens, Paninski, Sher, Litke,
  Chichilnisky and Simoncelli}]{Pillow2008}
\bibinfo{author}{Pillow, J.W.}, \bibinfo{author}{Shlens, J.},
  \bibinfo{author}{Paninski, L.}, \bibinfo{author}{Sher, A.},
  \bibinfo{author}{Litke, A.M.}, \bibinfo{author}{Chichilnisky, E.},
  \bibinfo{author}{Simoncelli, E.P.}, \bibinfo{year}{2008}.
\newblock \bibinfo{title}{Spatio-temporal correlations and visual signalling in
  a complete neuronal population}.
\newblock \bibinfo{journal}{Nature} \bibinfo{volume}{454},
  \bibinfo{pages}{995}.
\bibitem[{Posani et~al.(2017)Posani, Cocco, Je{\v{z}}ek and
  Monasson}]{Posani2017}
\bibinfo{author}{Posani, L.}, \bibinfo{author}{Cocco, S.},
  \bibinfo{author}{Je{\v{z}}ek, K.}, \bibinfo{author}{Monasson, R.},
  \bibinfo{year}{2017}.
\newblock \bibinfo{title}{Functional connectivity models for decoding of
  spatial representations from hippocampal {CA1} recordings}.
\newblock \bibinfo{journal}{J. Comput. Neurosci.} \bibinfo{volume}{43},
  \bibinfo{pages}{17--33}.
\bibitem[{Quinn et~al.(2011)Quinn, Coleman, Kiyavash and
  Hatsopoulos}]{Quinn2011}
\bibinfo{author}{Quinn, C.J.}, \bibinfo{author}{Coleman, T.P.},
  \bibinfo{author}{Kiyavash, N.}, \bibinfo{author}{Hatsopoulos, N.G.},
  \bibinfo{year}{2011}.
\newblock \bibinfo{title}{Estimating the directed information to infer causal
  relationships in ensemble neural spike train recordings}.
\newblock \bibinfo{journal}{J. Comput. Neurosci.} \bibinfo{volume}{30},
  \bibinfo{pages}{17--44}.
\bibitem[{Reid and Alonso(1995)}]{Reid1995}
\bibinfo{author}{Reid, R.C.}, \bibinfo{author}{Alonso, J.M.},
  \bibinfo{year}{1995}.
\newblock \bibinfo{title}{Specificity of monosynaptic connections from thalamus
  to visual cortex}.
\newblock \bibinfo{journal}{Nature} \bibinfo{volume}{378},
  \bibinfo{pages}{281}.
\bibitem[{Ren et~al.(2020)Ren, Ito, Hafizi, Beggs and traveledson}]{Ren2020}
\bibinfo{author}{Ren, N.}, \bibinfo{author}{Ito, S.}, \bibinfo{author}{Hafizi,
  H.}, \bibinfo{author}{Beggs, J.M.}, \bibinfo{author}{traveledson, I.H.},
  \bibinfo{year}{2020}.
\newblock \bibinfo{title}{Model-based detection of putative synaptic
  connections from spike recordings with latency and type constraints}.
\newblock \bibinfo{journal}{J. Neurophysiol.} \bibinfo{volume}{124},
  \bibinfo{pages}{1588--1604}.
\bibitem[{Sacerdote et~al.(2012)Sacerdote, Tamborrino and
  Zucca}]{Sacerdote2012}
\bibinfo{author}{Sacerdote, L.}, \bibinfo{author}{Tamborrino, M.},
  \bibinfo{author}{Zucca, C.}, \bibinfo{year}{2012}.
\newblock \bibinfo{title}{Detecting dependencies between spike trains of pairs
  of neurons through copulas}.
\newblock \bibinfo{journal}{Brain Res.} \bibinfo{volume}{1434},
  \bibinfo{pages}{243--256}.
\bibitem[{Scheffer et~al.(2020)Scheffer, Xu, Januszewski, Lu, Takemura,
  Hayworth, Huang, Shinomiya, Maitlin-Shepard, Berg et~al.}]{Scheffer2020}
\bibinfo{author}{Scheffer, L.K.}, \bibinfo{author}{Xu, C.S.},
  \bibinfo{author}{Januszewski, M.}, \bibinfo{author}{Lu, Z.},
  \bibinfo{author}{Takemura, S.y.}, \bibinfo{author}{Hayworth, K.J.},
  \bibinfo{author}{Huang, G.B.}, \bibinfo{author}{Shinomiya, K.},
  \bibinfo{author}{Maitlin-Shepard, J.}, \bibinfo{author}{Berg, S.}, et~al.,
  \bibinfo{year}{2020}.
\newblock \bibinfo{title}{A connectome and analysis of the adult drosophila
  central brain}.
\newblock \bibinfo{journal}{Elife} \bibinfo{volume}{9},
  \bibinfo{pages}{e57443}.
\bibitem[{Schmidt et~al.(2018)Schmidt, Bakker, Hilgetag, Diesmann and van
  Albada}]{schmidt2018multi}
\bibinfo{author}{Schmidt, M.}, \bibinfo{author}{Bakker, R.},
  \bibinfo{author}{Hilgetag, C.C.}, \bibinfo{author}{Diesmann, M.},
  \bibinfo{author}{van Albada, S.J.}, \bibinfo{year}{2018}.
\newblock \bibinfo{title}{Multi-scale account of the network structure of
  macaque visual cortex}.
\newblock \bibinfo{journal}{Brain Structure and Function}
  \bibinfo{volume}{223}, \bibinfo{pages}{1409--1435}.
\bibitem[{Schreiber(2000)}]{Schreiber2000}
\bibinfo{author}{Schreiber, T.}, \bibinfo{year}{2000}.
\newblock \bibinfo{title}{Measuring information transfer}.
\newblock \bibinfo{journal}{Phys. Rev. Lett.} \bibinfo{volume}{85},
  \bibinfo{pages}{461}.
\bibitem[{Schwindel et~al.(2014)Schwindel, Ali, McNaughton and
  Tatsuno}]{Schwindel2014}
\bibinfo{author}{Schwindel, C.D.}, \bibinfo{author}{Ali, K.},
  \bibinfo{author}{McNaughton, B.L.}, \bibinfo{author}{Tatsuno, M.},
  \bibinfo{year}{2014}.
\newblock \bibinfo{title}{Long-term recordings improve the detection of weak
  excitatory--excitatory connections in rat prefrontal cortex}.
\newblock \bibinfo{journal}{J. Neurosci.} \bibinfo{volume}{34},
  \bibinfo{pages}{5454--5467}.
\bibitem[{Seth(2007)}]{Seth2007}
\bibinfo{author}{Seth, A.}, \bibinfo{year}{2007}.
\newblock \bibinfo{title}{Granger causality}.
\newblock \bibinfo{journal}{Scholarpedia} \bibinfo{volume}{2},
  \bibinfo{pages}{1667}.
\bibitem[{Shao et~al.(2015)Shao, Huang, Shann, Yen, Tsai and Yen}]{Shao2015}
\bibinfo{author}{Shao, P.C.}, \bibinfo{author}{Huang, J.J.},
  \bibinfo{author}{Shann, W.C.}, \bibinfo{author}{Yen, C.T.},
  \bibinfo{author}{Tsai, M.L.}, \bibinfo{author}{Yen, C.C.},
  \bibinfo{year}{2015}.
\newblock \bibinfo{title}{Granger causality-based synaptic weights estimation
  for analyzing neuronal networks}.
\newblock \bibinfo{journal}{J. Comput. Neurosci.} \bibinfo{volume}{38},
  \bibinfo{pages}{483--497}.
\bibitem[{Shimono and Beggs(2015)}]{Shimono2015}
\bibinfo{author}{Shimono, M.}, \bibinfo{author}{Beggs, J.M.},
  \bibinfo{year}{2015}.
\newblock \bibinfo{title}{Functional clusters, hubs, and communities in the
  cortical microconnectome}.
\newblock \bibinfo{journal}{Cerebral Cortex} \bibinfo{volume}{25},
  \bibinfo{pages}{3743--3757}.
\bibitem[{Shorten et~al.(2021)Shorten, Spinney and Lizier}]{Shorten2021}
\bibinfo{author}{Shorten, D.P.}, \bibinfo{author}{Spinney, R.E.},
  \bibinfo{author}{Lizier, J.T.}, \bibinfo{year}{2021}.
\newblock \bibinfo{title}{Estimating transfer entropy in continuous time
  between neural spike trains or other event-based data}.
\newblock \bibinfo{journal}{PLoS Comput. Biol.} \bibinfo{volume}{17},
  \bibinfo{pages}{e1008054}.
\bibitem[{So et~al.(2012)So, Koralek, Ganguly, Gastpar and Carmena}]{So2012}
\bibinfo{author}{So, K.}, \bibinfo{author}{Koralek, A.C.},
  \bibinfo{author}{Ganguly, K.}, \bibinfo{author}{Gastpar, M.C.},
  \bibinfo{author}{Carmena, J.M.}, \bibinfo{year}{2012}.
\newblock \bibinfo{title}{Assessing functional connectivity of neural ensembles
  using directed information}.
\newblock \bibinfo{journal}{J. Neural Eng.} \bibinfo{volume}{9},
  \bibinfo{pages}{026004}.
\bibitem[{Song et~al.(2005)Song, Sj{\"o}str{\"o}m, Reigl, Nelson and
  Chklovskii}]{Song2005}
\bibinfo{author}{Song, S.}, \bibinfo{author}{Sj{\"o}str{\"o}m, P.J.},
  \bibinfo{author}{Reigl, M.}, \bibinfo{author}{Nelson, S.},
  \bibinfo{author}{Chklovskii, D.B.}, \bibinfo{year}{2005}.
\newblock \bibinfo{title}{Highly nonrandom features of synaptic connectivity in
  local cortical circuits}.
\newblock \bibinfo{journal}{PLoS Biol.} \bibinfo{volume}{3},
  \bibinfo{pages}{e68}.
\bibitem[{Spivak et~al.(2022)Spivak, Levi, Sloin, Someck and
  Stark}]{Spivak2022}
\bibinfo{author}{Spivak, L.}, \bibinfo{author}{Levi, A.},
  \bibinfo{author}{Sloin, H.E.}, \bibinfo{author}{Someck, S.},
  \bibinfo{author}{Stark, E.}, \bibinfo{year}{2022}.
\newblock \bibinfo{title}{Deconvolution improves the detection and
  quantification of spike transmission gain from spike trains}.
\newblock \bibinfo{journal}{Commun. Biol.} \bibinfo{volume}{5},
  \bibinfo{pages}{520}.
\bibitem[{Steinmetz et~al.(2021)Steinmetz, Aydin, Lebedeva, Okun, Pachitariu,
  Bauza, Beau, Bhagat, B{\"o}hm, Broux et~al.}]{Steinmetz2021}
\bibinfo{author}{Steinmetz, N.A.}, \bibinfo{author}{Aydin, C.},
  \bibinfo{author}{Lebedeva, A.}, \bibinfo{author}{Okun, M.},
  \bibinfo{author}{Pachitariu, M.}, \bibinfo{author}{Bauza, M.},
  \bibinfo{author}{Beau, M.}, \bibinfo{author}{Bhagat, J.},
  \bibinfo{author}{B{\"o}hm, C.}, \bibinfo{author}{Broux, M.}, et~al.,
  \bibinfo{year}{2021}.
\newblock \bibinfo{title}{Neuropixels 2.0: A miniaturized high-density probe
  for stable, long-term brain recordings}.
\newblock \bibinfo{journal}{Science} \bibinfo{volume}{372},
  \bibinfo{pages}{eabf4588}.
\bibitem[{Stetter et~al.(2012)Stetter, Battaglia, Soriano and
  Geisel}]{Stetter2012}
\bibinfo{author}{Stetter, O.}, \bibinfo{author}{Battaglia, D.},
  \bibinfo{author}{Soriano, J.}, \bibinfo{author}{Geisel, T.},
  \bibinfo{year}{2012}.
\newblock \bibinfo{title}{Model-free reconstruction of excitatory neuronal
  connectivity from calcium imaging signals}.
\newblock \bibinfo{journal}{PLoS Comput. Biol.} \bibinfo{volume}{8},
  \bibinfo{pages}{e1002653}.
\bibitem[{Stevenson(2023)}]{Stevenson2023}
\bibinfo{author}{Stevenson, I.H.}, \bibinfo{year}{2023}.
\newblock \bibinfo{title}{Circumstantial evidence and explanatory models for
  synapses in large-scale spike recordings}.
\newblock \bibinfo{journal}{Neurons Behav. Data Anal. Theory}
  \bibinfo{volume}{December}, \bibinfo{pages}{1--30}.
\bibitem[{Stevenson and Kording(2011)}]{Stevenson2011a}
\bibinfo{author}{Stevenson, I.H.}, \bibinfo{author}{Kording, K.P.},
  \bibinfo{year}{2011}.
\newblock \bibinfo{title}{How advances in neural recording affect data
  analysis}.
\newblock \bibinfo{journal}{Nat. Neurosci.} \bibinfo{volume}{14},
  \bibinfo{pages}{139}.
\bibitem[{Stevenson et~al.(2008a)Stevenson, Rebesco, Hatsopoulos, Haga, Miller
  and Kording}]{Stevenson2008b}
\bibinfo{author}{Stevenson, I.H.}, \bibinfo{author}{Rebesco, J.M.},
  \bibinfo{author}{Hatsopoulos, N.G.}, \bibinfo{author}{Haga, Z.},
  \bibinfo{author}{Miller, L.E.}, \bibinfo{author}{Kording, K.P.},
  \bibinfo{year}{2008}a.
\newblock \bibinfo{title}{Bayesian inference of functional connectivity and
  network structure from spikes}.
\newblock \bibinfo{journal}{IEEE Trans. Neural Syst. Rehabil. Eng.}
  \bibinfo{volume}{17}, \bibinfo{pages}{203--213}.
\bibitem[{Stevenson et~al.(2008b)Stevenson, Rebesco, Miller and
  K{\"o}rding}]{Stevenson2008a}
\bibinfo{author}{Stevenson, I.H.}, \bibinfo{author}{Rebesco, J.M.},
  \bibinfo{author}{Miller, L.E.}, \bibinfo{author}{K{\"o}rding, K.P.},
  \bibinfo{year}{2008}b.
\newblock \bibinfo{title}{Inferring functional connections between neurons}.
\newblock \bibinfo{journal}{Curr. Opin. Neurobiol.} \bibinfo{volume}{18},
  \bibinfo{pages}{582--588}.
\bibitem[{Swadlow(1995)}]{Swadlow1995}
\bibinfo{author}{Swadlow, H.A.}, \bibinfo{year}{1995}.
\newblock \bibinfo{title}{Influence of vpm afferents on putative inhibitory
  interneurons in s1 of the awake rabbit: evidence from cross-correlation,
  microstimulation, and latencies to peripheral sensory stimulation}.
\newblock \bibinfo{journal}{J. Neurophysiol.} \bibinfo{volume}{73},
  \bibinfo{pages}{1584--1599}.
\bibitem[{Terada et~al.(2020)Terada, Obuchi, Isomura and
  Kabashima}]{Terada2020}
\bibinfo{author}{Terada, Y.}, \bibinfo{author}{Obuchi, T.},
  \bibinfo{author}{Isomura, T.}, \bibinfo{author}{Kabashima, Y.},
  \bibinfo{year}{2020}.
\newblock \bibinfo{title}{Inferring neuronal couplings from spiking data using
  a systematic procedure with a statistical criterion}.
\newblock \bibinfo{journal}{Neural Comput.} \bibinfo{volume}{32},
  \bibinfo{pages}{2187--2211}.
\bibitem[{Thomson et~al.(2002)Thomson, West, Wang and Bannister}]{Thomson2002}
\bibinfo{author}{Thomson, A.M.}, \bibinfo{author}{West, D.C.},
  \bibinfo{author}{Wang, Y.}, \bibinfo{author}{Bannister, A.P.},
  \bibinfo{year}{2002}.
\newblock \bibinfo{title}{Synaptic connections and small circuits involving
  excitatory and inhibitory neurons in layers 2--5 of adult rat and cat
  neocortex: triple intracellular recordings and biocytin labelling in vitro}.
\newblock \bibinfo{journal}{Cerebral Cortex} \bibinfo{volume}{12},
  \bibinfo{pages}{936--953}.
\bibitem[{Toyama et~al.(1981)Toyama, Kimura and Tanaka}]{Toyama1981}
\bibinfo{author}{Toyama, K.}, \bibinfo{author}{Kimura, M.},
  \bibinfo{author}{Tanaka, K.}, \bibinfo{year}{1981}.
\newblock \bibinfo{title}{Organization of cat visual cortex as investigated by
  cross-correlation technique.}
\newblock \bibinfo{journal}{J. Neurophysiol.} \bibinfo{volume}{46},
  \bibinfo{pages}{202--214}.
\bibitem[{Truccolo et~al.(2005)Truccolo, Eden, Fellows, Donoghue and
  Brown}]{Truccolo2005}
\bibinfo{author}{Truccolo, W.}, \bibinfo{author}{Eden, U.T.},
  \bibinfo{author}{Fellows, M.R.}, \bibinfo{author}{Donoghue, J.P.},
  \bibinfo{author}{Brown, E.N.}, \bibinfo{year}{2005}.
\newblock \bibinfo{title}{A point process framework for relating neural spiking
  activity to spiking history, neural ensemble, and extrinsic covariate
  effects}.
\newblock \bibinfo{journal}{J. Neurophysiol.} \bibinfo{volume}{93},
  \bibinfo{pages}{1074--1089}.
\bibitem[{Tsubo and Shinomoto(2023)}]{tsubo2023non}
\bibinfo{author}{Tsubo, Y.}, \bibinfo{author}{Shinomoto, S.},
  \bibinfo{year}{2023}.
\newblock \bibinfo{title}{Non-differentiable activity in the brain}.
\newblock \bibinfo{journal}{bioRxiv} , \bibinfo{pages}{2023--06}.
\bibitem[{Vicente et~al.(2011)Vicente, Wibral, Lindner and Pipa}]{Vicente2011}
\bibinfo{author}{Vicente, R.}, \bibinfo{author}{Wibral, M.},
  \bibinfo{author}{Lindner, M.}, \bibinfo{author}{Pipa, G.},
  \bibinfo{year}{2011}.
\newblock \bibinfo{title}{Transfer entropy―a model-free measure of effective
  connectivity for the neurosciences}.
\newblock \bibinfo{journal}{J. Comput. Neurosci.} \bibinfo{volume}{30},
  \bibinfo{pages}{45--67}.
\bibitem[{Xia et~al.(2019)Xia, Liu, Bai, Zheng and Tian}]{Xia2019}
\bibinfo{author}{Xia, M.}, \bibinfo{author}{Liu, T.}, \bibinfo{author}{Bai,
  W.}, \bibinfo{author}{Zheng, X.}, \bibinfo{author}{Tian, X.},
  \bibinfo{year}{2019}.
\newblock \bibinfo{title}{Information transmission in hpc-pfc network for
  spatial working memory in rat}.
\newblock \bibinfo{journal}{Behav. Brain Res.} \bibinfo{volume}{356},
  \bibinfo{pages}{170--178}.
\bibitem[{Yoshimura et~al.(2005)Yoshimura, Dantzker and
  Callaway}]{Yoshimura2005}
\bibinfo{author}{Yoshimura, Y.}, \bibinfo{author}{Dantzker, J.L.},
  \bibinfo{author}{Callaway, E.M.}, \bibinfo{year}{2005}.
\newblock \bibinfo{title}{Excitatory cortical neurons form fine-scale
  functional networks}.
\newblock \bibinfo{journal}{Nature} \bibinfo{volume}{433},
  \bibinfo{pages}{868}.
\bibitem[{Yuan and Lin(2006)}]{Yuan2006}
\bibinfo{author}{Yuan, M.}, \bibinfo{author}{Lin, Y.}, \bibinfo{year}{2006}.
\newblock \bibinfo{title}{Model selection and estimation in regression with
  grouped variables}.
\newblock \bibinfo{journal}{J. R. Stat. Soc. Series B: Stat. Methodol.}
  \bibinfo{volume}{68}, \bibinfo{pages}{49--67}.
\bibitem[{Zaytsev et~al.(2015)Zaytsev, Morrison and Deger}]{Zaytsev2015}
\bibinfo{author}{Zaytsev, Y.V.}, \bibinfo{author}{Morrison, A.},
  \bibinfo{author}{Deger, M.}, \bibinfo{year}{2015}.
\newblock \bibinfo{title}{Reconstruction of recurrent synaptic connectivity of
  thousands of neurons from simulated spiking activity}.
\newblock \bibinfo{journal}{J. Comput. Neurosci.} \bibinfo{volume}{39},
  \bibinfo{pages}{77--103}.

\end{thebibliography}
%





\end{document}